\documentclass[aps,prb,twocolumn,showpacs,superscriptaddress,groupedaddress]{revtex4-1}
\pdfoutput=1
\usepackage{graphicx}  % for figures
\usepackage{xcolor}
\usepackage{bm}        % for math
\usepackage{amssymb}   % for math
\usepackage{amsmath}   % for align environment
\usepackage{color}
\usepackage{natbib}
\usepackage{cancel}
\usepackage[caption=false,labelformat=simple]{subfig}  % to correctly label figure parts 
\usepackage[normalem]{ulem} % for strikethrough
\definecolor{orange}{rgb}{0.7,0.2,0}
\definecolor{darkgreen}{rgb}{0,0.6,0}
\let\oldsqrt\sqrt  % renames \sqrt as \oldsqrt
\def\sqrt{\mathpalette\DHLhksqrt} % defines the new \sqrt in terms of the old one
\def\DHLhksqrt#1#2{%
\setbox0=\hbox{$#1\oldsqrt{#2\,}$}\dimen0=\ht0
\advance\dimen0-0.2\ht0
\setbox2=\hbox{\vrule height\ht0 depth -\dimen0}%
{\box0\lower0.4pt\box2}}
\begin{document}
\title{Correlated charge transport in bilinear tunnel junction arrays}
\author{Kelly A. Walker}\email[]{kelly.walker@rmit.edu.au}
\author{Jared H. Cole}
\affiliation{Chemical and Quantum Physics, School of Applied Sciences, RMIT University, Melbourne, 3001, Australia}
%\date{\today}
%----------------------------------------------------------
%-----ABSTRACT-----
%----------------------------------------------------------
\begin{abstract}
We study theoretically the nature of correlations in space and time of the current in a one-dimensional bilinear array of tunnel junctions in the normal conduction limit, using the kinetic Monte Carlo method. The bilinear array consists of two parallel rows of tunnel junctions,  capacitively coupled in a ladder configuration. The electrostatic potential landscape and the charge-charge interaction length both depend on the circuit capacitances, which in turn influence transport and charge correlations in the array. We observe the formation of stationary charge states when only one rail is voltage biased. When a symmetric bias is applied to both rails, the site at which the positive and negative charge carriers recombine can drift throughout the array. We also calculate charge densities and auto- and cross-correlation functions.
\end{abstract}
\pacs{03.75.Lm, 73.63.-b, 73.23.Hk}
\maketitle
%----------------------------------------------------------
%-----INTRODUCTION-----
\section{\label{sec:intro}Introduction}
%----------------------------------------------------------
Tunnel junction arrays are interesting devices as they straddle the boundary between discrete systems such as quantum dots and single-electron transistors (SETs) and more continuous systems such as one-dimensional (1D) quantum wires. Arrays of tunnel junctions are also a candidate device for a quantum definition of the ampere\cite{Averin:1989}. Such devices display correlated transport properties due to the inherent electrostatic interactions of excitations moving through the array\cite{Grabert:1991}, which can extend across many junctions. These interactions can manifest as quasibound charge pairs, solitons or the formation of a Wigner lattice, where the latter is central to their use in metrology. These correlated charges have been observed directly in real time as periodic, discrete spikes of current, each associated with one electron\cite{Bylander:2005}.

\begin{figure}[th!]
\centering
\subfloat{
\includegraphics[width=0.44\textwidth]{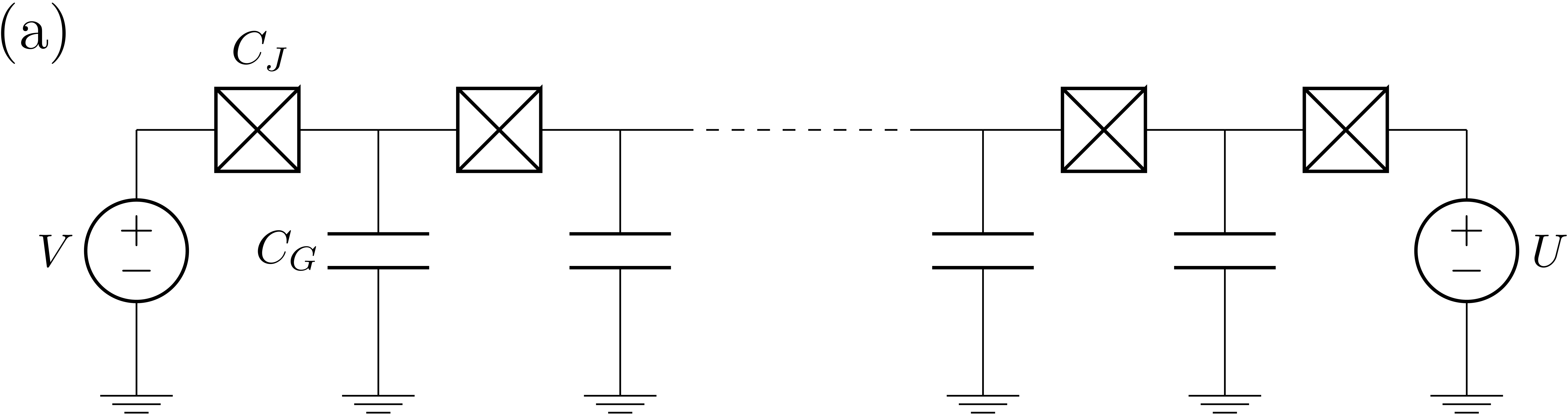}   \label{fig:linear_array}}
\vspace{4mm}
\subfloat{
\includegraphics[width=0.44\textwidth]{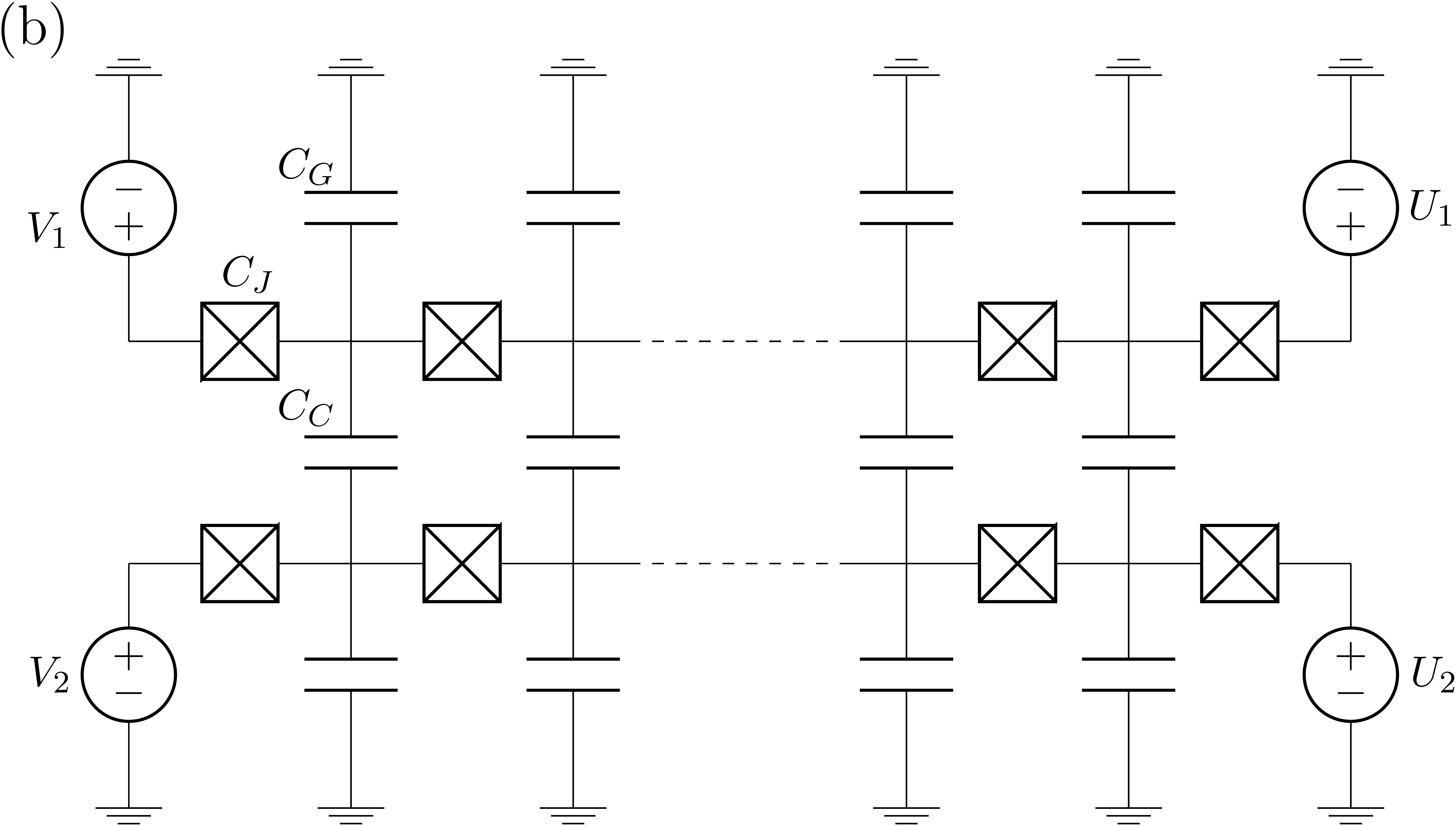}   \label{fig:bilinear_array}}
\caption{(a) Linear array circuit diagram consisting of $N-1$ tunnel junctions. Voltage can be applied at either or both ends and each junction has a junction capacitance $C_J$. Each island is capacitively coupled to a ground plane through a ground capacitor $C_G$. (b) Bilinear array circuit diagram consisting of \emph{two} capacitively coupled linear tunnel junction arrays. Each island is capacitively coupled to the corresponding parallel island in the opposite rail by a capacitance $C_C$.
} 
\label{fig:schematic}
\end{figure}

Bilinear arrays of tunnel junctions consist of two parallel linear arrays (rails) of tunnel junctions capacitively coupled by a capacitance $C_{C}$, see Fig.~\ref{fig:bilinear_array}. Charge transport through bilinear arrays can be carried by effective excitons \cite{Averin:1991}, whereby electrons flow through one rail and holes flow in the other rail. This phenomenon is due to the long-range Coulomb potential of individual charge carriers, which enables the rails to exchange momentum and energy and forms the basis of the current drag effect\cite{Hansch:1983,Shimada:2000}. Even in the limit of zero tunneling between rails, Coulomb interactions between rails are sufficient to open a charge gap and generate correlations between charges in opposite rails. Correlated transport has been measured in bilinear tunnel junction arrays via correlated noise measurements of the current\cite{Shimada:2000}, however it has not been measured at the level of single charges.

In this paper we study the nature of the correlations in space and time of the current in a biased bilinear array in the normal conduction limit. We begin by discussing the transport properties in a 1D \emph{linear} array consisting of $N=50$ identical islands separated by $N-1$ identical tunnel junctions (Fig.~\ref{fig:linear_array}) to provide a comparison to the 1D bilinear array, which consists of \emph{two} $N=50$ capacitively coupled tunnel junction arrays (Fig.~\ref{fig:bilinear_array}).  As the probability for cotunneling events spanning the system dramatically decreases with increasing resistance $R$, cotunneling is neglected and charge transport is dominated by incoherent single-electron transitions. 

As an electron propagates along the array, it not only raises the potential of its island, but also the surrounding islands, thereby preventing other electrons from tunneling into the area. This exponential repulsive interaction $U$ experienced by a pair of charges on sites $m$ and $n$ is
\begin{equation}
U\propto e^{-|m-n|/\Lambda}
\end{equation}  
where the length $\Lambda$ depends on the circuit capacitances---the junction $C_J$ and the gate capacitance $C_G$---and can be approximated by\cite{Delsing:1992,Hu:1994}
\begin{equation}
\Lambda\approx\sqrt{\frac{C_J}{C_G}}
\end{equation}
This interaction length characterizes the spatial separation between charges and therefore the properties of the correlated transport. 

We used the stochastic kinetic Monte Carlo (KMC) method\cite{Likharev:1989,Bakhvalov:1989} to model the charge dynamics in nonlinear time. By exploring stochastic sequences of transitions via the Gillespie algorithm\cite{Gillespie:1977}, KMC generates highly accurate and efficient simulations of the temporal evolution of the system that replicate experimental observations. Following the usual orthodox theory\cite{Wasshuber:2001,Ingold:1992,Averin:1991bk,Grabert:1992bk,Likharev:1985}, the transition rate associated with transition $i\rightarrow{j}$ is governed by
\begin{equation}
\Gamma_{i\rightarrow{j}}=\frac{\Delta E}{q_e^{2}R_T}\frac{1}{e^{\Delta{E}/k_BT}+1} \label{eq: rate}  
\end{equation}
where $\Delta{E}$ is the change in energy of the transition, $q_e$ is the elementary charge, $R_T$ is the junction tunnel resistance, $k_B$ is the Boltzmann constant and $T$ is the electron temperature. The time between tunneling events $\Delta{t}$ (i.e.~the time the system spends in a specific charge configuration $i$) represents a single KMC time step. The KMC algorithm is repeated, keeping track of all quantities, until good statistics are obtained. We use at least $10^6$ Monte Carlo steps and ensure the system has equilibrated before collecting statistics.

All simulations presented here assume an electron temperature of $T=30$ mK. At this temperature the total capacitance of each island is sufficiently small so that the charging energy $E_C=q_e^2/2C_\Sigma$ of the islands is much greater than the energy of thermal fluctuations $k_{B}T$. We assume that $R_T$ is much greater than the quantum (Klitzing) resistance $R_{T}\gg{R}_{K}=h/q_e^{2}$. For these simulations $R_T=10^6$ $\Omega$, but the exact value is unimportant as this simply rescales the time between Monte Carlo steps. The ground and junction capacitances are $C_{G}=2$ aF and $C_{J}=50$ aF, respectively, which gives our interaction length $\Lambda=5$. We can also reasonably assume that the energy relaxation is significantly faster than all other dynamics, allowing the electrons to be described by a Fermi distribution. These conditions ensure that the system is always in a well-defined charge state, therefore the orthodox theory of single-electron tunneling\cite{Averin:1991bk,Grabert:1992bk,Likharev:1985} applies. 

Beginning with an empty array, we equilibrate the circuit (evolve until a stable charge configuration is reached) before collecting statistics. In an experimental array, uncontrolled charged impurities (background charges) induce an additional random offset charge on every island. Such background charges affect both threshold voltages $V_{\mathrm{(th)}}$ and the soliton flow\cite{Melsen:1997,Bylander:2007dt,Johansson:2000}. Arrays with a short soliton length are more sensitive to these irregularities in the potential from island to island. Here we neglect background charges as we specifically consider correlations in relatively long arrays. 
%----------------------------------------------------------
%-----LINEAR ARRAY-----
\section{\label{sec:lin}Linear array}
%----------------------------------------------------------
We begin by summarizing the important results for correlated charge transport in a symmetrically biased ($V=-U=\Delta{V/2}$) linear array of $N=50$ islands, see Fig.~\ref{fig:linear_array}. This investigation provides a comparison to that of the bilinear array.

Using the $N\times{N}$ capacitance matrix,
\begin{equation}
\mathbf{C}_{mn} = \begin{pmatrix}
C_G+2C_J & -C_J & 0 &\hdots \\
-C_J & C_G+2C_J & -C_J &\ddots \\
0 & -C_J  & \ddots & \ddots & \\
\vdots  &  \ddots&\ddots&\\ 
&&&&&& \end{pmatrix} 
\end{equation}
and the method proposed by Devoret\cite{Devoret:1997} and recently summarized by Fay \emph{et al.}\cite{Fay:2011}, we describe the system by the Hamiltonian,
\begin{align}
\mathcal H=\sum_{n,m}\bigg[&\frac{1}{2}{\mathbf{C}^{-1}_{mn}}Q_nQ_m \notag\\
&+\delta_{n,1}\mathbf{C}^{-1}_{mn} Q_m C_J V +\delta_{n,N}\mathbf{C}^{-1}_{mn}Q_mC_JU\bigg]
\end{align} 
where $Q_{n}$ is the charge on the $n$th site and $\delta$ is the Kronecker delta function. Using the Hamiltonian we can compute the energy of an arbitrary charge configuration. 

Taking the analytic form of the Hamiltonian, we can understand the origin of correlated transport. Using an analytic inverse of the capacitance matrix\cite{Hu:1994}, we derive the interaction energy between two charges in the linear array. Assuming an infinite array (i.e.~$N\rightarrow\infty$) and setting $V=U=0$, the interaction energy between two charges $Q_m$ and $Q_n$ close to the center of the array ($m,n\approx{N/2}$) is given by
\begin{align}
U(Q_m,Q_n)=&\frac{{Q^{2}_m}}{4C_J}\Bigg(\frac{1}{\sinh\lambda}\Bigg)+\frac{{Q^{2}_n}}{4C_J}\Bigg(\frac{1}{\sinh\lambda}\Bigg)\notag\\
&+\frac{Q_mQ_n}{2C_J}\Bigg(\frac{e^{-|n-m|/\Lambda}}{\sinh\lambda}\Bigg)
\end{align}
where $\lambda=1/\Lambda$. In the limit of long interaction length ($\Lambda\gg1$) this gives
\begin{align}
U(Q_m,Q_n)=&\frac{{Q^{2}_m}}{4\sqrt{C_JC_G}}+\frac{{Q^{2}_n}}{4\sqrt{C_JC_G}}\notag\\
&+\frac{Q_mQ_n}{2\sqrt{C_JC_G}}\Big(e^{-|n-m|/\Lambda}\Big)
\end{align}
This expression is composed of two charging energy terms plus an interaction energy term.

\begin{figure}[th!]
\centering
\includegraphics[width=0.47\textwidth]{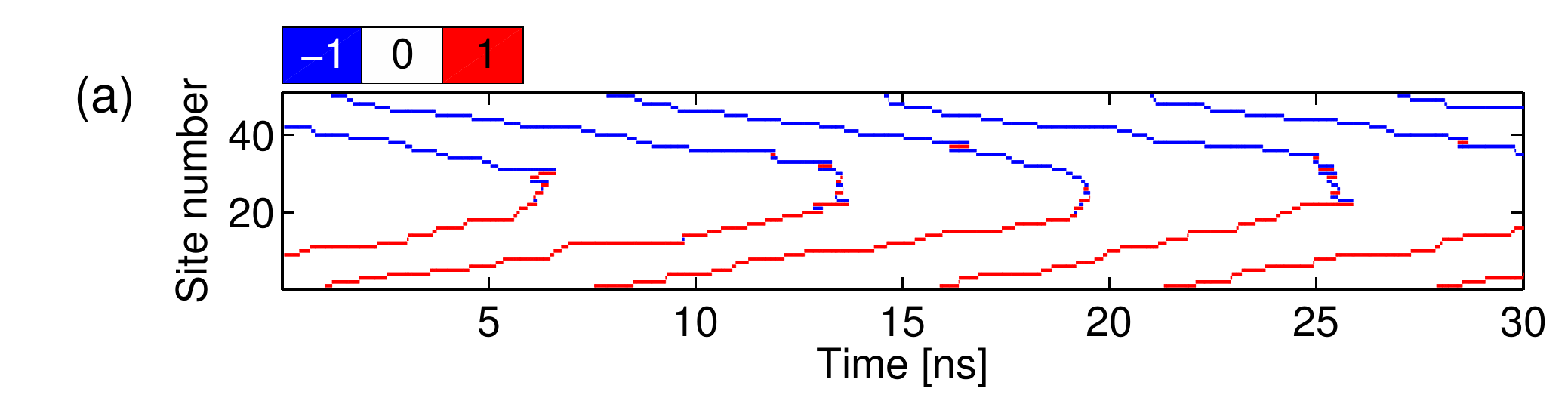}   % \label{fig:linear_charge_s-t}
\includegraphics[width=0.47\textwidth]{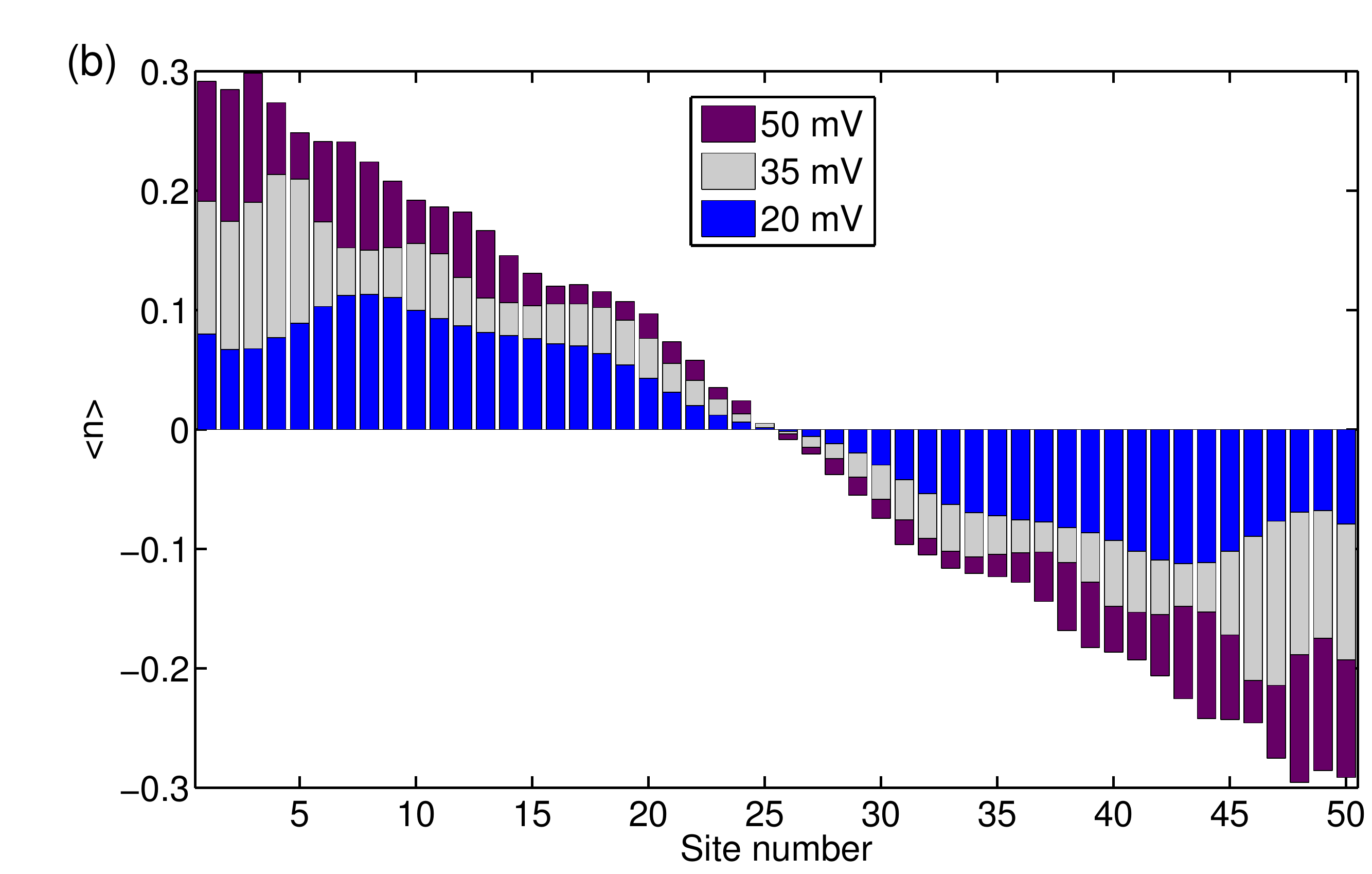}     %\label{fig:linear_chargedist}
\caption{(Color online) (a) Charge occupancy diagram as a function of time and position within the linear array at $\Delta{V}=20$ mV. Correlated electrons (red) and holes (blue) flow in opposite directions. White represents the zero charge state. Many dipole states form near the recombination site. (b) Average charge distribution as a function of position. For low voltages there are periodic oscillations in the charge distribution.  This can be attributed to the (temporally) correlated transport seen in (a) and the effect of the boundaries of the array.
} 
\label{fig:linear_charge}
\end{figure}

A similar analysis can be used to determine the threshold voltage where conduction begins. To find $V_{\mathrm{(th)}}$ we set the voltage equal to the energy of a charge on the first island\cite{Hu:1994},
\begin{equation}
V_{\mathrm{(th)}}=\frac{q_e}{2C_J(e^{\lambda}-1)}\label{thresvlinear} 
\end{equation}
In the limit of long interaction length ($\Lambda\gg1$) this gives
\begin{equation}
V_{\mathrm{(th)}}\approx\frac{q_e}{2\sqrt{C_JC_G}}
\end{equation}
where for our model, Eq.~\ref{thresvlinear} gives $\Delta{V_{\mathrm{(th)}}}=14.5$ mV.

The charge occupancy diagram in Fig.~\ref{fig:linear_charge}(a) shows that both charge carriers---electrons and holes---exhibit strong time-correlated charge transport at every site $n$. Due to the symmetric potential $\Delta{V}$, charges of the same sign cannot tunnel from one end of the array to the other. Instead electrons recombine in the middle of the array with holes tunneling in the opposite direction, resulting in a net flow of current. Furthermore, the correlations are more spread out at the center of the rail as the potential drop between sites is not even throughout the length of the array, i.e.~there is not a uniform drop in potential per site.

\begin{figure}[th!]
\includegraphics[width=0.483\textwidth]{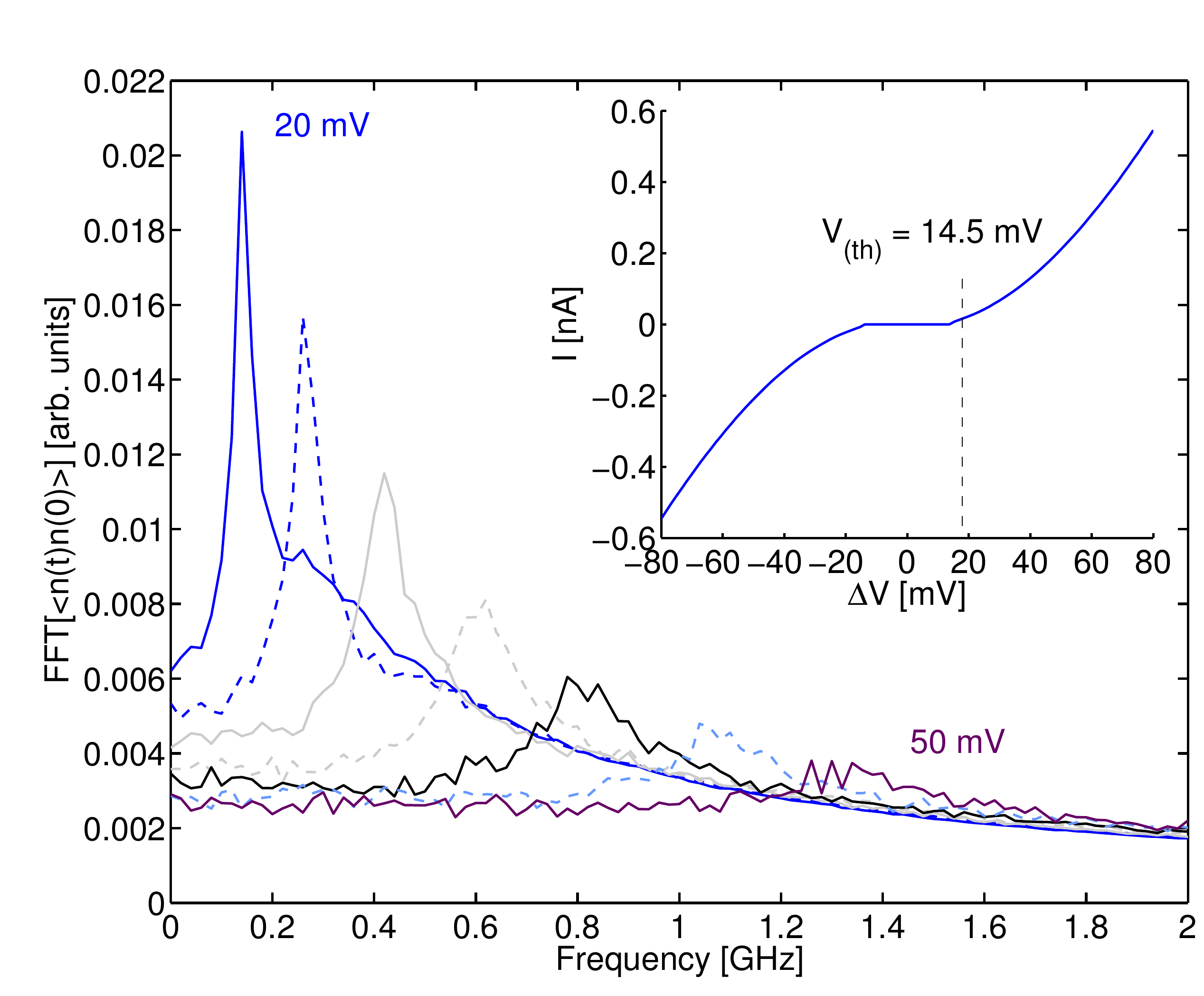}
\caption{(Color online) Spectral response of the linear array charge-charge correlation function measured at  $n=5$, for increasing voltage in steps of 5 mV. For increasing voltage, the peak frequency increases but the peak amplitude diminishes due to higher average charge densities which reduce the formation of correlated charge states. At low voltages, the peak frequency $f_p$ corresponds to an effective charge carrier of precisely $q_e$, ie. $f_p=I/q_e$. Inset: Current-voltage characteristics measured at the first junction and characterized by a Coulomb blockade---a zero-current state for bias $|\Delta{V}|$ below the characteristic $\Delta{V_{\mathrm{(th)}}}=14.5$ mV, Eq.~\ref{thresvlinear}.
}
\label{fig:corr_linear}
\end{figure}

A dipole state is created within the array when a charge induces a neighboring charge of opposite polarity. The creation and recombination of such dipole states increases with $\Lambda$ as the charges can exert a greater influence on their nearest neighbor sites and the creation of a hole is more energetically favorable\cite{Homfeld:2011}. The energy required to create a dipole within the array (relative to the energy of a single charge) is
\begin{equation}
U(1,-1)/U(1,0)=1/2\Lambda
\end{equation}
therefore the energy required to create a dipole decreases as the separation between charges ($\Lambda$) increases.

We can already see evidence of correlated transport in the average charge distribution within the array, Fig.~\ref{fig:linear_charge}(b). At low $\Delta{V}$, the charge distribution exhibits periodic peaks, a signature of correlated transport.

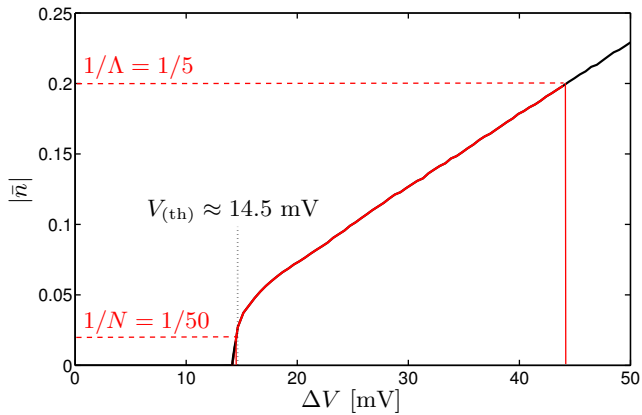
\begin{figure}[tbp]
\begin{center}
\setlength{\unitlength}{1cm}
\begin{picture}(7.4,6)
\put(-0.8,0.5){\includegraphics[width=8.7\unitlength]{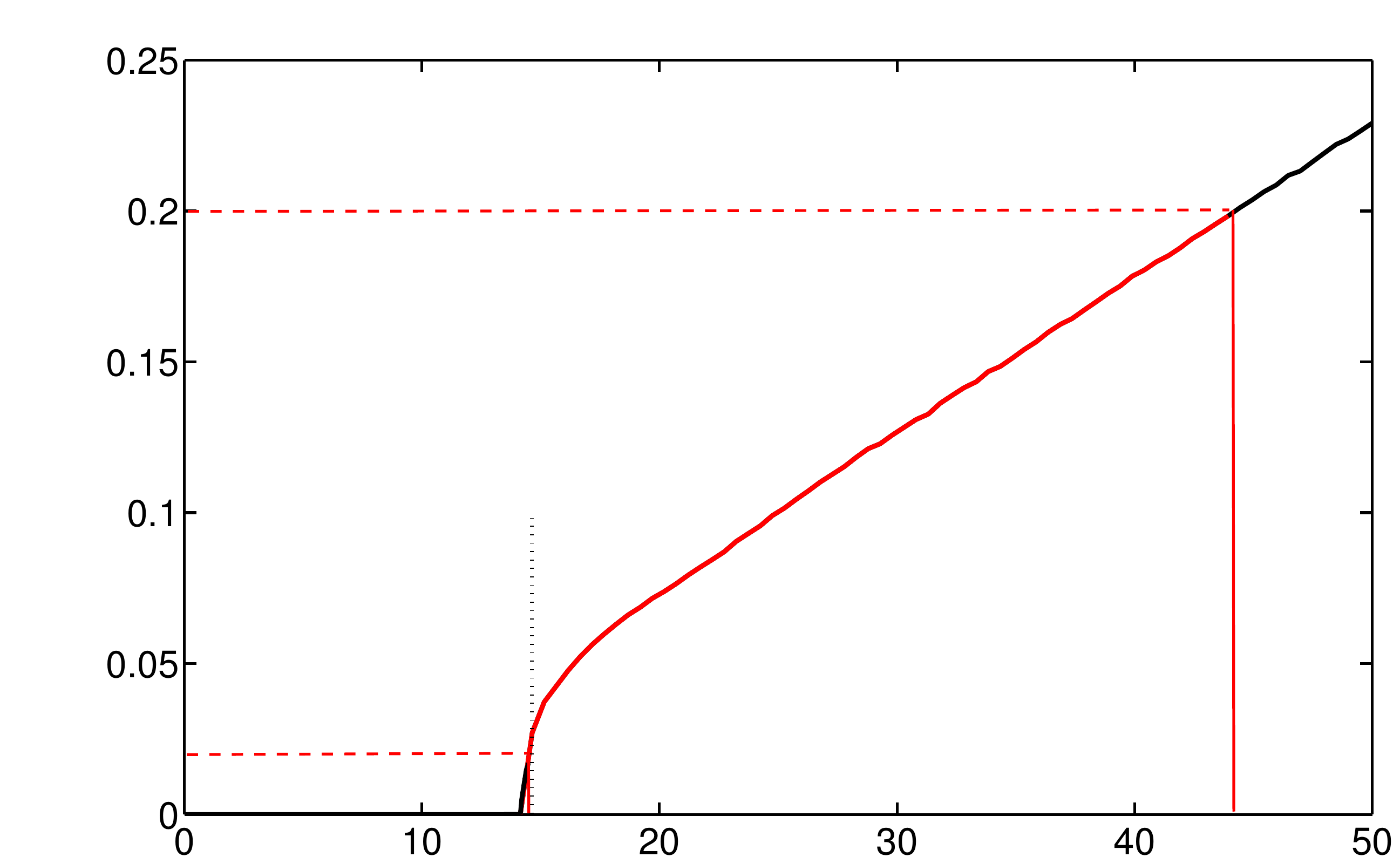}}
\put(-0.54,2.95){\rotatebox{90}{$|\bar{n}|$}}
\put(3.35,0.25){$\Delta{V}$ [mV]}
\put(1.3,2.8){$V_{\mathrm{(th)}}\approx14.5$ mV} 
\put(0.45,4.7){\textcolor{red}{$1/\Lambda=1/5$}}
\put(0.45,1.3){\textcolor{red}{$1/N=1/50$}}
\end{picture}
\caption{(Color online) Average charge density within the linear array as a function of $\Delta{V}$. For $\Delta{V}\gtrsim44$ mV, on average the charges are separated by less than $\Lambda$. For higher voltages, the charge separation is reduced and this ultimately leads to the breakdown of correlated transport (see Fig.~\ref{fig:corr_linear}).
}
\label{fig:avechargeden}
\end{center}
\end{figure}

To investigate these correlations, we compute the charge-charge correlation spectrum $\mathcal{F}[\langle Q_n(\tau)Q_m(0)\rangle]$, see Fig.~\ref{fig:corr_linear}. In principle we can calculate the correlations between charges directly from the KMC output\cite{Likharev:1989}. However, we find that performing a linear sampling of the data (we use a bandwidth BW = 20 GHz throughout) and then taking the fast Fourier transform (FFT) provides a more efficient method for calculating the spectrum of the charge-charge correlations. Note that while state-of-the-art charge detectors have a lower BW than the 20 GHz used in our simulations, the parameters in our model ($C_J$, $R_T$) could be optimized experimentally to produce spectra within the detection window of the chosen detector.

The correlation functions show clear and distinct peaks, indicating strong correlations in the transport carriers in these junction arrays. The correlations are robust due to the large correlation length ($\Lambda=5$). This is consistent with the charge occupancy diagram Fig.~\ref{fig:linear_charge}(a), where we also saw strong correlations between charge carriers---even the annihilation of the electron-hole pairs occurs periodically. As $\Delta{V}$ increases and the applied bias exceeds the force of Coulomb repulsion, correlations begin to decay. We see this as the gradual flattening out and disappearance of the correlation peak.

This breakdown of correlated transport can be understood from Fig.~\ref{fig:avechargeden}, where we show the average charge density for the entire array. Correlated charge transport begins at the onset of conduction $\Delta{V_{\mathrm{(th)}}}$ when at least one charge is present in the array. For $\Lambda=5$, one could expect that the optimal separation between charges is $\sim5$ sites, i.e.~this is the period. Therefore there are optimally $\sim10$ charges in the linear array at any one time. However, increasing $\Delta{V}$ injects more charge carriers into the array, increasing the charge density (thereby reducing the average charge separation). As charges are pushed closer and closer together, the applied voltage overwhelms the Coulomb force responsible for periodic separation of the charges. When charges can no longer maintain their well-defined positions with respect to one another, correlations are destroyed and so the peaks are suppressed. We observe transport through the array to become predominately uncorrelated when $\Delta{V}\gtrsim50$ mV. 

The strength of the correlations also varies with position within the array, where correlations become progressively weaker towards the center where opposing carriers recombine. This is consistent with Fig.~\ref{fig:linear_charge}(b), where the charge distribution is approximately zero at the center of the array.
%----------------------------------------------------------
%-----BILINEAR ARRAY-----
\section{\label{sec:bi}Bilinear array}
%----------------------------------------------------------
We now turn our attention to a \emph{bilinear} array, consisting of two $N=50$ linear arrays of islands, see Fig.~\ref{fig:bilinear_array}. The two arrays are capacitively coupled by $C_C$ (i.e.~not by tunnel junctions), so charges cannot tunnel between the two rails. 

We consider three different biasing regimes: symmetric single-rail, symmetric dual-rail and antisymmetric dual-rail (escalator) biasing. In the symmetric single-rail case, a symmetric potential bias $\Delta{V}$ is only applied to the upper rail, i.e.~$\Delta V_{1}/2=V_{1}=-U_{1}$. In the symmetric dual-rail biasing regime, a symmetric potential $\Delta{V}$ is applied to both rails, i.e.~$\Delta V_{1}=\Delta V_{2}$. Finally, in the bias regime we term escalator biasing, a symmetric bias is applied to both rails, but with opposite sign, i.e.~$\Delta V_{1}=-\Delta V_{2}$. Two different coupling strengths are also investigated: weak $C_C=C_G$ and strong $C_C=5\times{C_G}=C_J/5$.

We use the method discussed in Sec.~\ref{sec:lin} to construct the capacitance matrix,
\begin{widetext}
\begin{equation}
\mathbf{C}_{mn}=\begin{pmatrix}
C_G+2C_J+C_C & -C_J&0 & 0 &\hdots&-C_C&0&\hdots \\
-C_J & C_G+2C_J+C_C &-C_J& 0 &0&\hdots&-C_C&0 \\
0  & -C_J& C_G+2C_J+C_C & -C_J&0 &0 &\hdots&-C_C \\
0  & 0& -C_J&\ddots&\ddots&\ddots&\ddots&\ddots\\
\vdots&0&0&\ddots&\ddots&\ddots&\ddots&\ddots\\
-C_C&\vdots&0&\ddots&\ddots &\ddots&\ddots&\ddots\\
0&-C_C&\vdots&\ddots&\ddots&\ddots &\ddots &\ddots \\
\vdots&0&-C_C&\ddots&\ddots&\ddots &\ddots&\ddots  
\end{pmatrix} 
\end{equation}
\end{widetext}
Note that the bilinear array capacitance matrix is essentially a double copy of the linear array except for the inclusion of the coupling terms. Similarly, the Hamiltonian is given by
\begin{align}
\mathcal H= \sum_{n,m}\bigg[&\frac{1}{2}{\mathbf{C}^{-1}_{mn}}Q_nQ_m \notag\\
& +\delta_{n,1}\mathbf{C}^{-1}_{mn} Q_mC_JV_1 + \delta_{n,N+1}\mathbf{C}^{-1}_{mn}Q_mC_JV_2  \notag\\
&+\delta_{n,N}\mathbf{C}^{-1}_{mn}Q_mC_JU_1 +  \delta_{n,2N}\mathbf{C}^{-1}_{mn}Q_mC_JU_2\bigg] 
\end{align}
where we can now apply a bias voltage at either end of both rails.

There are two cases for the interaction energy in the bilinear array, either where charges $m$ and $n$ are in the \emph{same} rail or where charges $m$ and $n$ are in \emph{different} rails. Similarly to the linear case, we can use the analytic inversion of the bilinear capacitance matrix\cite{Hu:1996} to obtain expressions for the interaction energy between two charges $Q_m$ and $Q_n$,
\begin{align}
U_{\epsilon}(Q_m,Q_n)=&\frac{{Q^{2}_m}}{8C_J}\bigg(\frac{1}{\sinh\lambda_{+}}+\frac{1}{\sinh\lambda_{-}}\bigg) \nonumber\\
&+\frac{{Q^{2}_n}}{8C_J}\bigg(\frac{1}{\sinh\lambda_{+}}+\frac{1}{\sinh\lambda_{-}}\bigg) \nonumber\\
&+\frac{Q_mQ_n}{4C_J}\bigg(\frac{e^{-\lambda_{+}|m-n|}}{\sinh\lambda_{+}}+\epsilon\frac{e^{-\lambda_{-}|m-n|}}{\sinh\lambda_{-}}\bigg)
\end{align}
where $\epsilon =\pm1$ corresponds to charges within the same rail or different rails, respectively, and
where $\lambda_\pm$ is defined by 
\begin{equation}
2\cosh\lambda_\pm=2+\frac{C_G}{C_J}+\frac{C_C}{C_J}(1\mp1) \label{lambda}
\end{equation}

Again, similarly to the linear case, we calculate $\Delta{V_{\mathrm{(th)}}}$ for fixed $\Delta{V_2}$ for a symmetric bias, assuming positive voltage,
\begin{align}
\Delta{V_{1_{\mathrm{(th)}}}} =\left\{
\begin{aligned}
&\frac{q_eA_{11}}{C_J(1-A_{11})}+\frac{\Delta{V_2}B_{11}}{(1-A_{11})},  &\Delta{V_2}>0 \\
&\frac{q_eA_{11}}{C_J(1-A_{11})}, &\Delta{V_2}\leq0
\label{thresv}
\end{aligned}
\right.
\end{align}
where $A_{11}=\frac{1}{2}[e^{-\lambda_+}+e^{-\lambda_-}]$ and $B_{11}=\frac{1}{2}[e^{-\lambda_+}-e^{-\lambda_-}]$ for large $N$. The threshold voltage is similar for $\Delta{V_2}$, where the conduction threshold is whichever of the two ($\Delta{V_{1_{\mathrm{(th)}}}}$ or $\Delta{V_{2_{\mathrm{(th)}}}}$) is lower.
\begin{figure}[th!]
\centering
\includegraphics[width=0.477\textwidth]{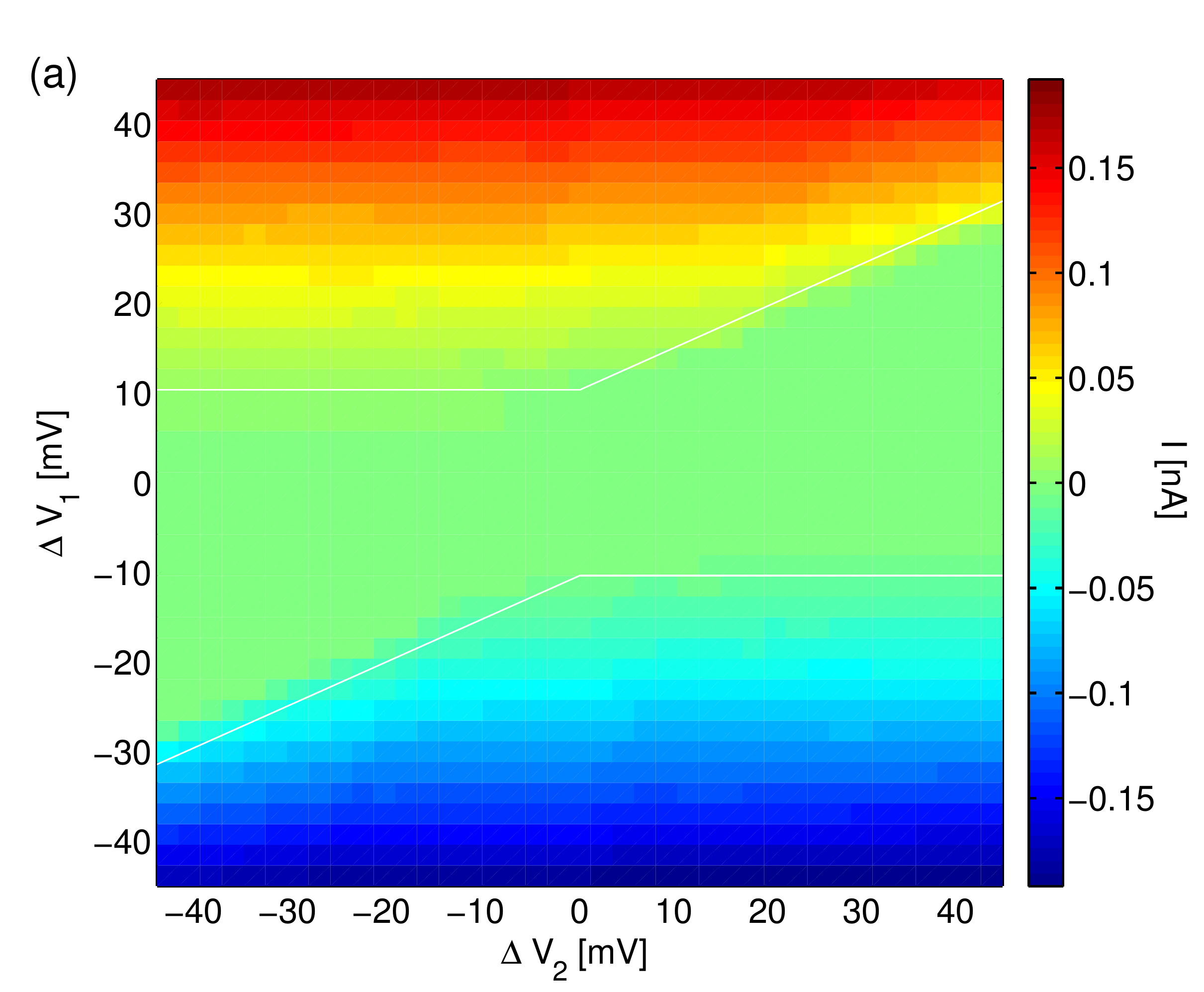}\label{fig:pcolor_IV_upper}%\subfloat[]{   % to include subfigure labeling.

\includegraphics[width=0.477\textwidth]{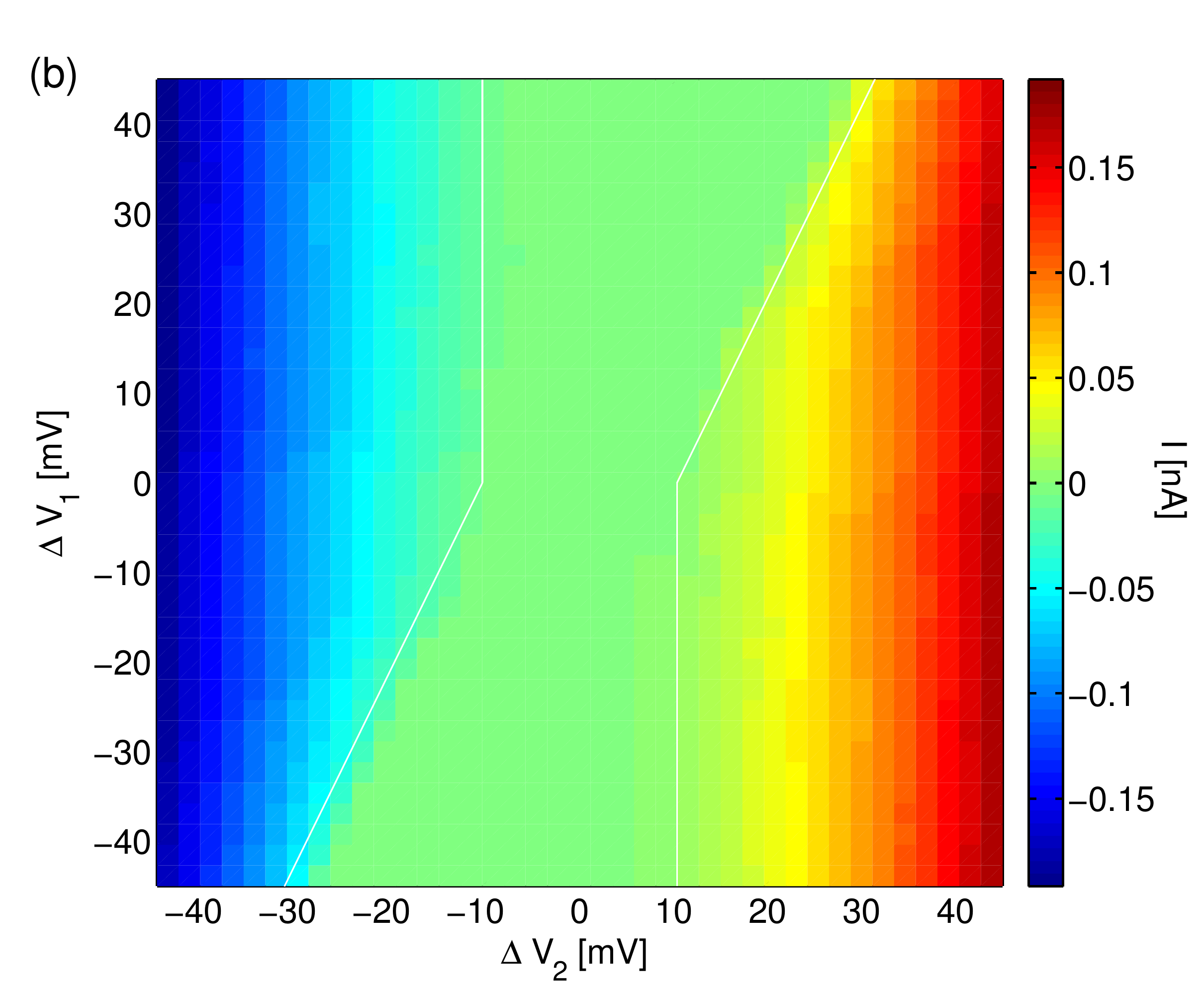}\label{fig:pcolor_IV_lower}
\caption{(Color online) Current-voltage characteristics of the (a) upper and (b) lower rails with weak $C_C$. $\Delta{V_1}$ ($\Delta {V_2}$) is applied to the upper (lower) rail. White contour lines show the analytical conduction voltage $\Delta{V_{\mathrm{(th)}}}$.
}
\label{fig:pcolorIV}
\end{figure}

In Fig.~\ref{fig:pcolorIV}, we plot the current-voltage characteristics for weak $C_C$ and show the threshold voltages for the upper and lower rails, calculated by Eq.~\ref{thresv}. The region between these lines represents the Coulomb blockade state, wherein there is zero charge flow. Outside these lines (i.e.~$\geq{\Delta{V}_{\mathrm{(th)}}}$), the rail conducts, although conduction begins slowly. The current-voltage characteristics for strong $C_C$ are similar except for the increased Coulomb gap along the line corresponding to escalator bias, due to the dipole injection effect, discussed in Sec.~\ref{sec:sc}.
%----------------------------------------------------------
%-----STATIC STATES-----
\section{\label{sec:staticstates}Static states: Single-rail biasing}
\begin{figure}[th!]
\centering
\includegraphics[width=0.47\textwidth]{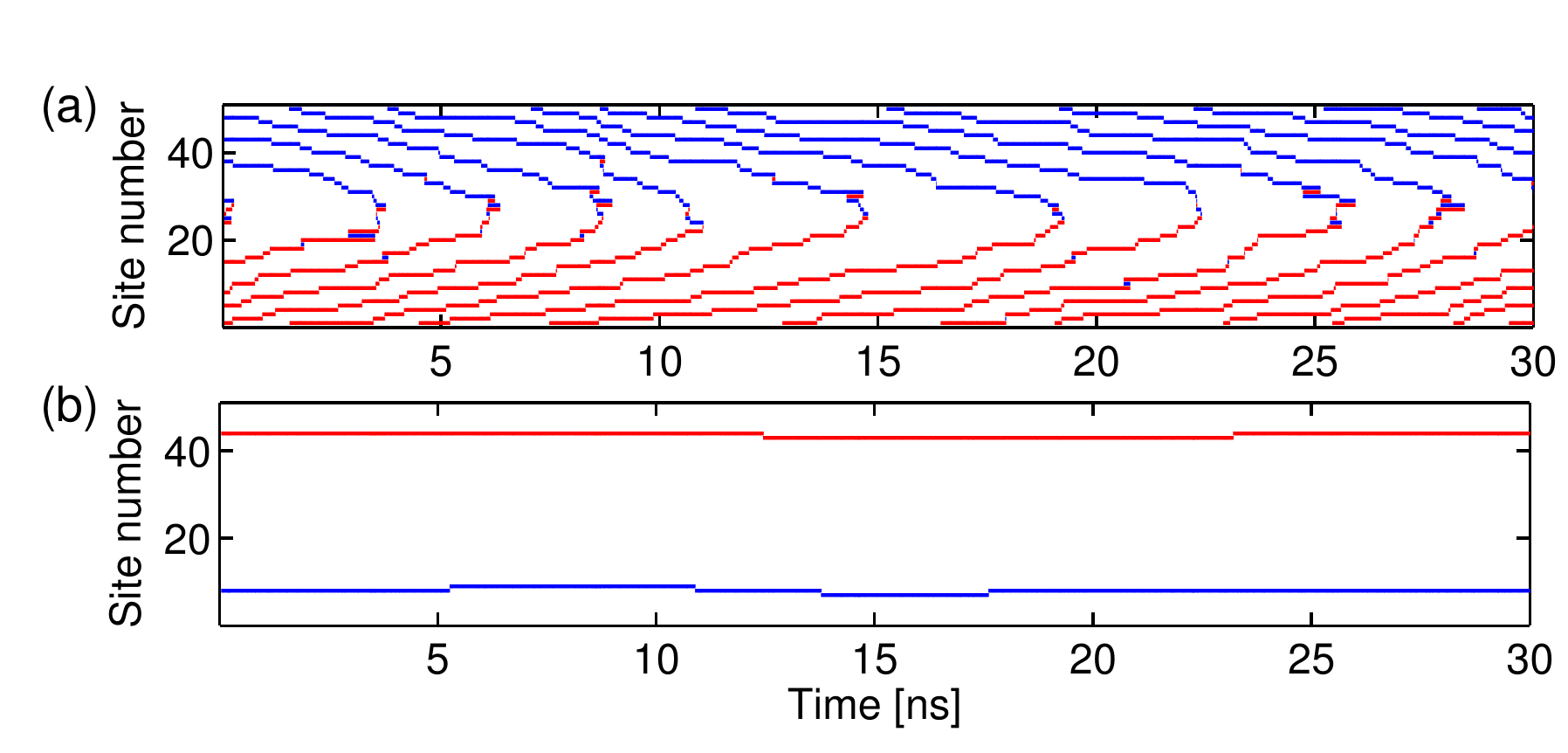}
\caption{(Color online) Charge occupancy diagrams for symmetric single-rail bias $\Delta V_{1}=25$ mV within the (a) driven and (b) undriven rails with weak $C_C$. In the undriven rail, we see stationary charge states, i.e.~the rail does not conduct at this value of $\Delta V_{1}$.
}
\label{fig:charge_s-t_weakSSRB}
\end{figure}

%----------------------------------------------------------
When a symmetric single-rail bias ($\Delta{V_{1}}\neq0$, $\Delta{V_2}=0$) is applied and the rails are decoupled (i.e.~$C_C=0$), as expected no parasitic current or static charge states are induced in the undriven rail because the rails are independent, whereas weak $C_C$ (i.e.~$C_C=C_G$) allows the undriven rail to weakly feel the potential of the driven rail. While it is not sufficient to produce parasitic current, static charge states are created in the undriven rail, whereby holes enter from the left and electrons enter from the right, Fig.~\ref{fig:charge_s-t_weakSSRB}(b). These charges penetrate a finite distance into the array but do not result in a net current, but rather in stationary charge states (or charge polarization). While these states are not temporally correlated, they do exhibit spatial correlation. As a result, in the \emph{single-rail bias} regime, we do not observe moving charge correlations in the driven and undriven rails simultaneously because the energy differences of the two rails are too great and interrail correlations are always destroyed before a drag current is observed.
\begin{figure}[th!]
\centering
\includegraphics[width=0.47\textwidth]{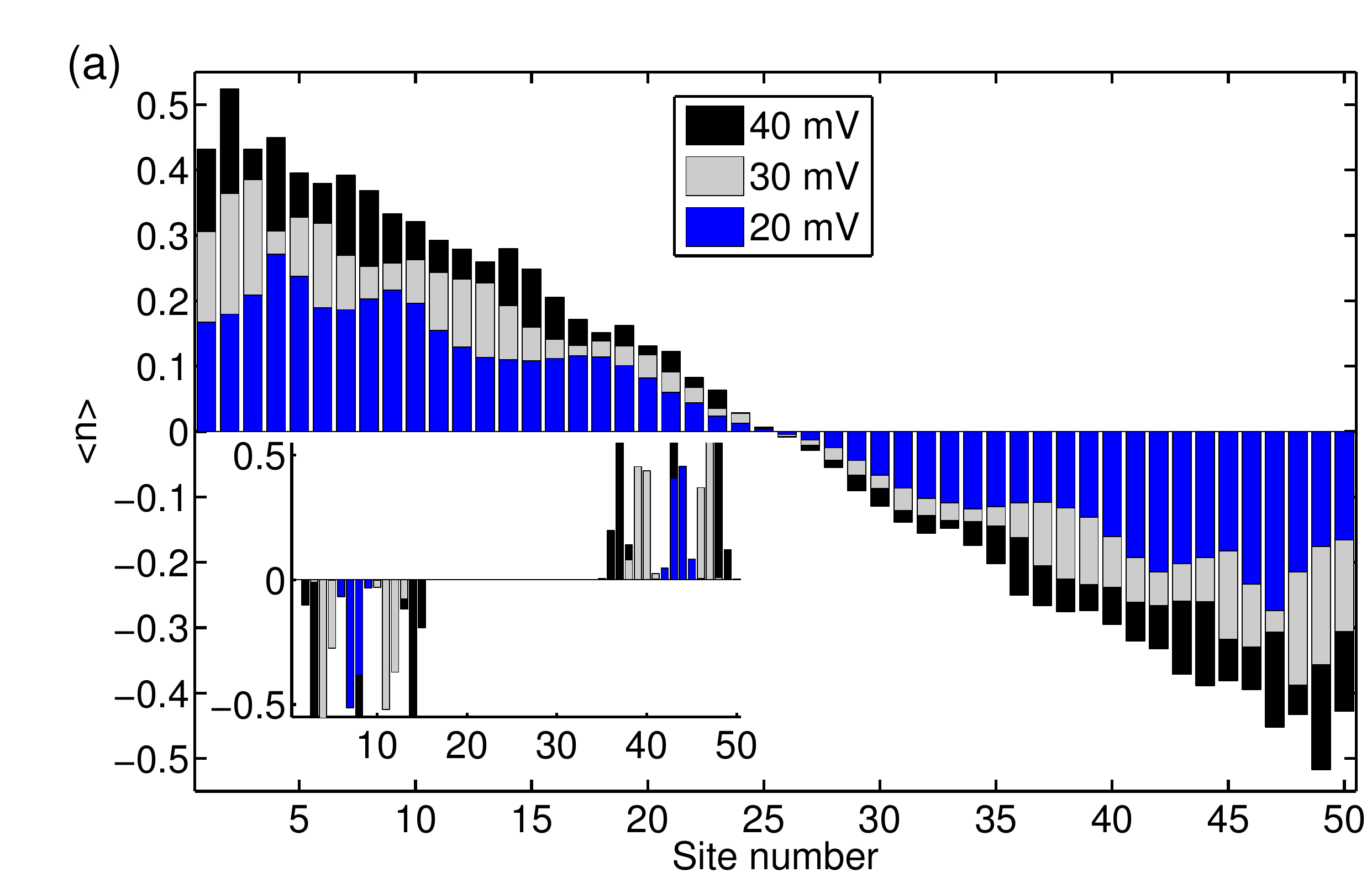}%\label{fig:chargedist_sym_upper}

\includegraphics[width=0.47\textwidth]{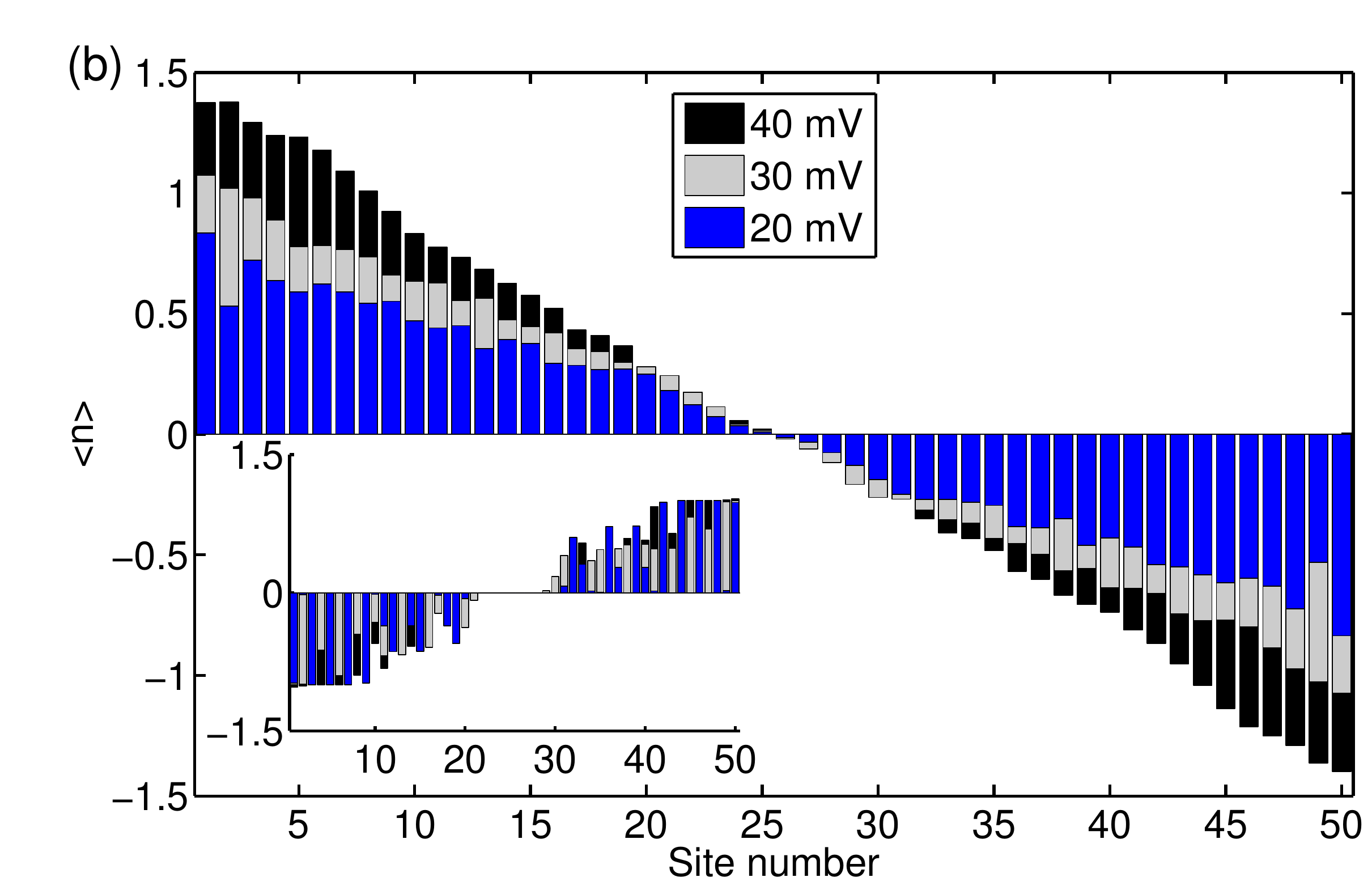}%\label{fig:chargedist_sym_strong}
\caption{(Color online) Average charge distribution per site for symmetric single-rail bias. For (a) weak $C_C$, the peaks in the driven rail charge distribution at low $\Delta V_{1}$ are indicative of correlated transport. Inset: Stationary states form in the undriven rail. (b) Strong $C_C$ reduces correlations in the driven rail and (inset) induces a greater number of static states in the undriven rail.
}
\label{fig:chargedist_sym}
\end{figure}

Fig.~\ref{fig:chargedist_sym} shows the average charge distribution within the array for both weak and strong $C_C$. The peaks in the driven rail charge distribution for weak $C_C$ at low $\Delta V_{1}$, Fig.~\ref{fig:chargedist_sym}(a), are a signature of correlated transport, whereas stationary states form in the undriven rail. As $\Delta V_{1}$ is increased, the site occupancy in the driven rail becomes much greater than one and correlations are suppressed.
\begin{figure}[th!]
\centering
\includegraphics[width=0.47\textwidth]{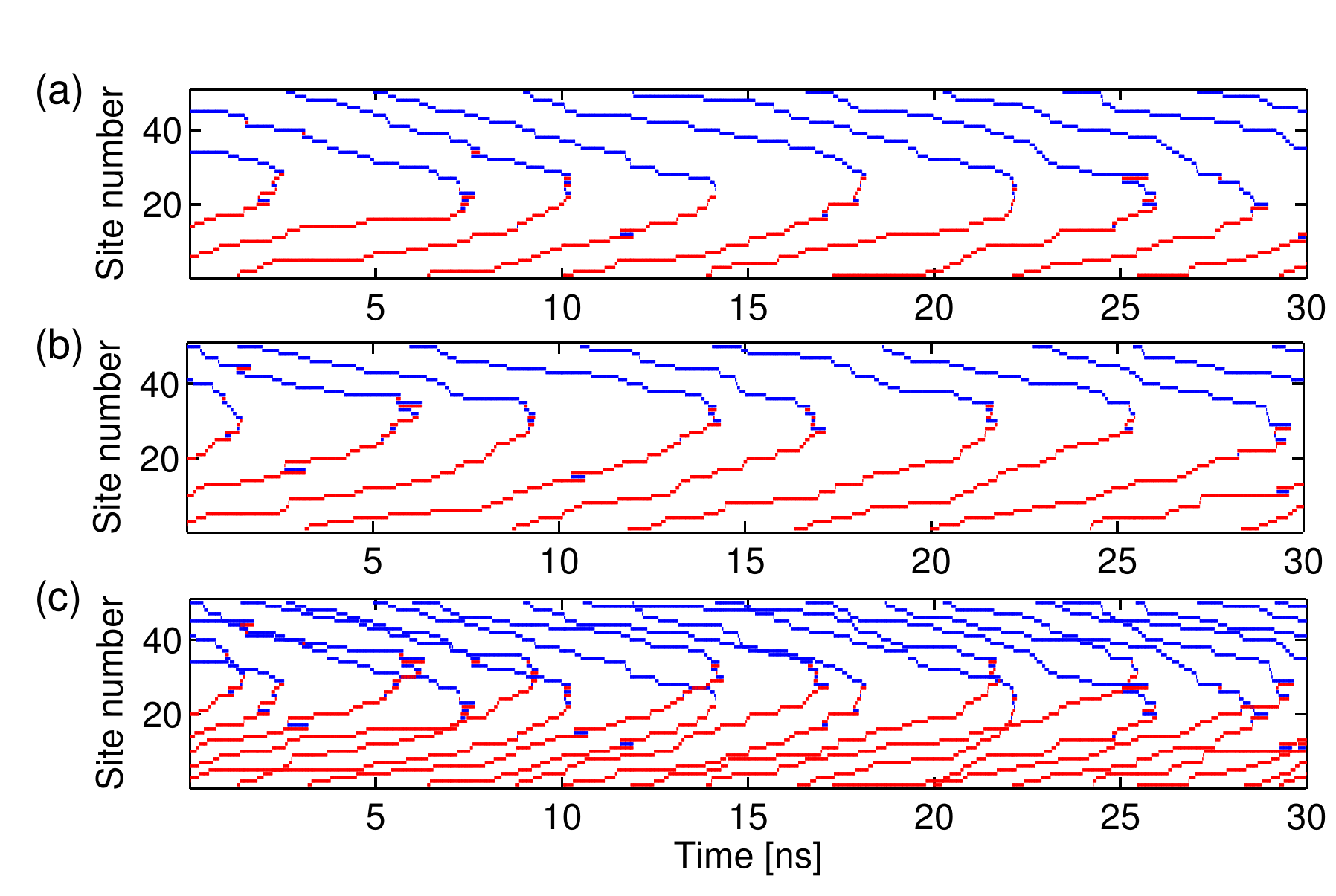}
\caption{(Color online) Charge occupancy diagrams for symmetric dual-rail bias $\Delta{V}=25$ mV within the (a) upper and (b) lower rails with weak $C_C$. (c) shows (a) superimposed on (b). The currents are out of phase with one another, evidence of synchronized correlations.
}
\label{fig:charge_s-t_symdualrail_weak}
\end{figure}

One typically expects larger $C_C$ to lead to current drag effects, however increasing $C_C$ suppresses charge correlation in the driven rail, Fig.~\ref{fig:chargedist_sym}(b), at lower $\Delta{V_1}$ and creates a greater number of static states in the undriven rail, as discussed above. We therefore see again that \emph{single-rail biasing} cannot induce a parasitic current while \emph{also} displaying correlated transport.
%----------------------------------------------------------
%-----SYNCHRONIZED CORRELATIONS-----
\section{\label{sec:sc}Synchronized correlations: Dual-rail biasing}
\begin{figure}[th!]
\centering
\includegraphics[width=0.47\textwidth]{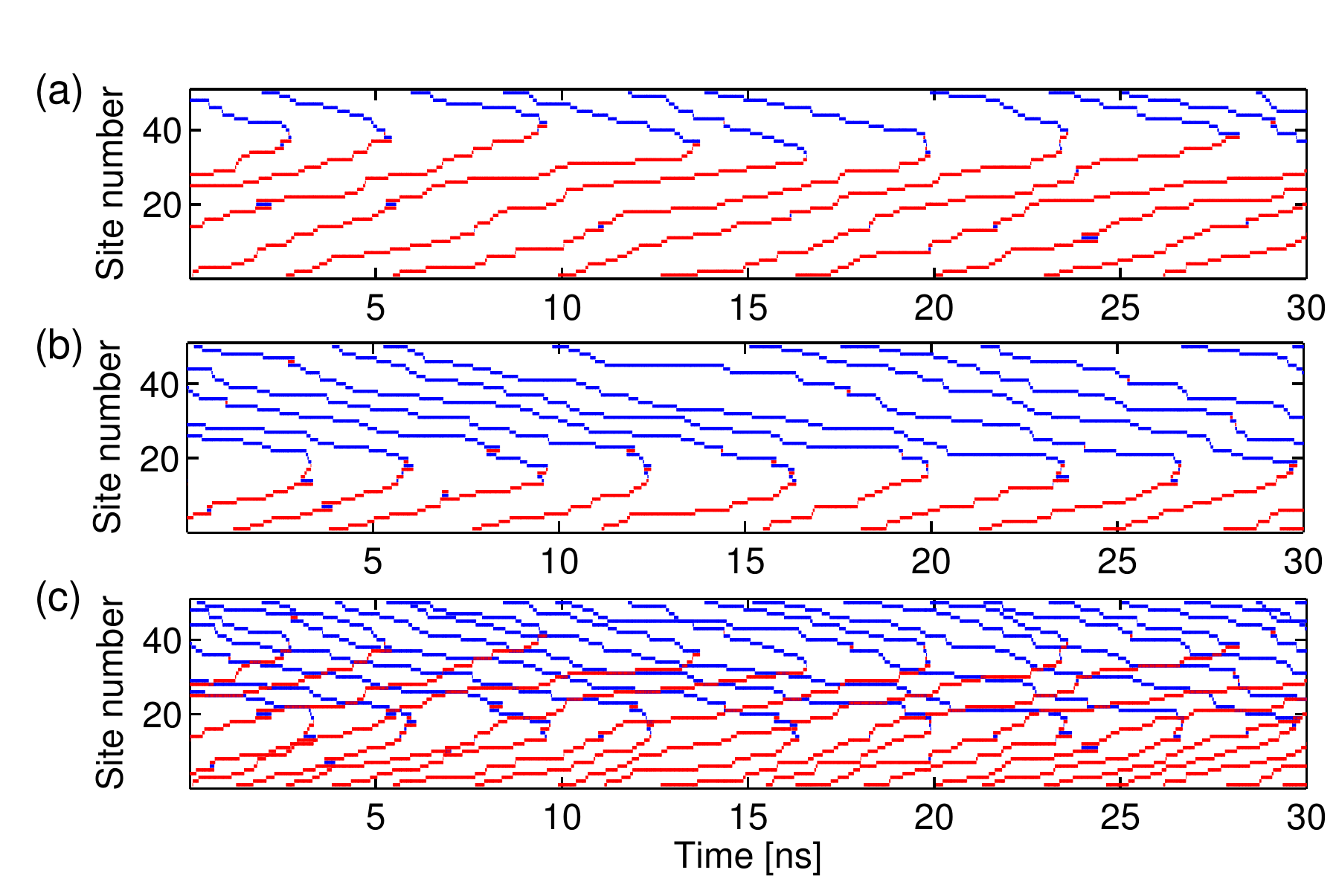}
\caption{(Color online) Charge occupancy diagrams for symmetric dual-rail bias $\Delta V=25$ mV within the (a) upper and (b) lower rails with strong $C_C$. The upper and lower rail recombination sites move in unison, creating an unequal number of electrons (holes) in each rail. In (c), where (a) is superimposed on (b), we see that the currents are out of phase with one another, suggesting that the currents are anticorrelated.
}
\label{fig:charge_s-t_symdualrail_strong}
\end{figure}

%----------------------------------------------------------
We now turn our attention to equally driving both rails, looking for synchronized correlations. First, we consider a symmetric dual-rail bias ($\Delta{V_1}=\Delta{V_2}$). The space-time diagrams show strong correlated transport in both rails for both weak (Fig.~\ref{fig:charge_s-t_symdualrail_weak}) and strong (Fig.~\ref{fig:charge_s-t_symdualrail_strong}) $C_C$. In addition, when the upper and lower rail charge occupancy diagrams are superimposed, we see that the correlations are synchronized. Rather than seeing the current correlations precisely overlap, we see that the correlations tend to be out of phase with one another, signifying charge locking, which results in strong (anti-) correlations \textit{between} rails (i.e.~cross correlations).

\begin{figure}[th!]
\includegraphics[width=0.483\textwidth]{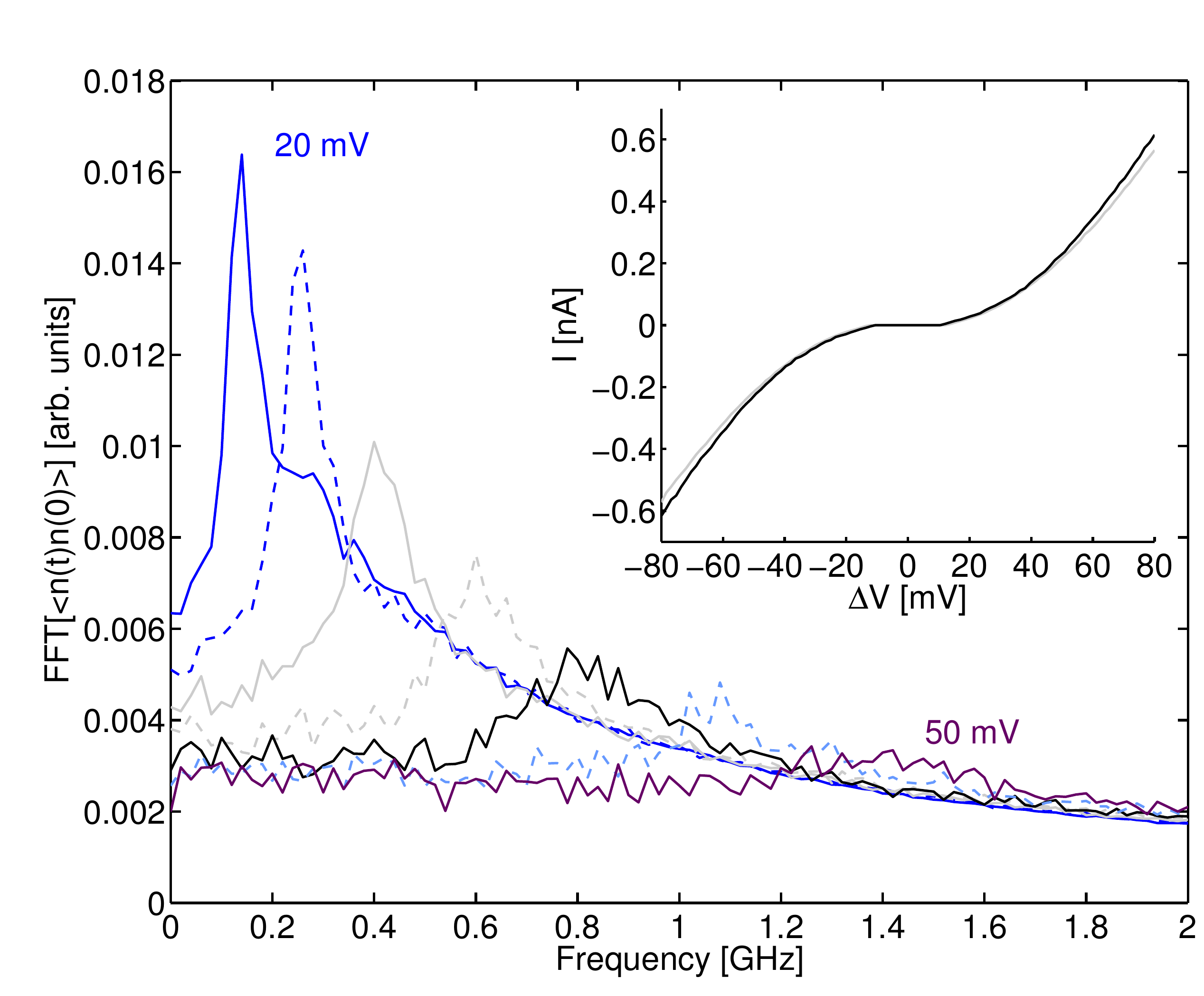}
\caption{(Color online) Spectral response of the charge-charge correlation function for symmetric dual-rail bias with weak $C_C$ calculated at $n=5$, for increasing voltage in steps of 5 mV. Similarly to the linear case, as voltage increases, the peak frequency increases linearly with increasing current, but higher voltage weakens the correlations. Inset: Current-voltage characteristics of the upper rail measured at the first junction. Lower rail characteristics are identical for this bias type. Strong $C_C$ (black) induces slightly larger currents and smaller Coulomb gaps than weak $C_C$ (gray).
}
\label{fig:corr_symdualrail} 
\end{figure}

The autocorrelation functions in Fig.~\ref{fig:corr_symdualrail} show clear and distinct peaks, indicating strong correlations within a rail. As the applied voltage begins to dominate over Coulomb repulsion, the robust correlation peak gradually decays and flattens out, indicating that correlation has been lost. Similarly to the linear case, the strength of the correlations varies at different positions within the array. The charge carriers become increasingly correlated as they tunnel towards the center of the array away from edge effects. The autocorrelation functions are also near identical for symmetric sites (upper and lower rails).
As seen in the linear case, the correlations are weakest in the center of the array due to the reduced average charge occupancy.
\begin{figure}[th!]
\centering
\includegraphics[width=0.47\textwidth]{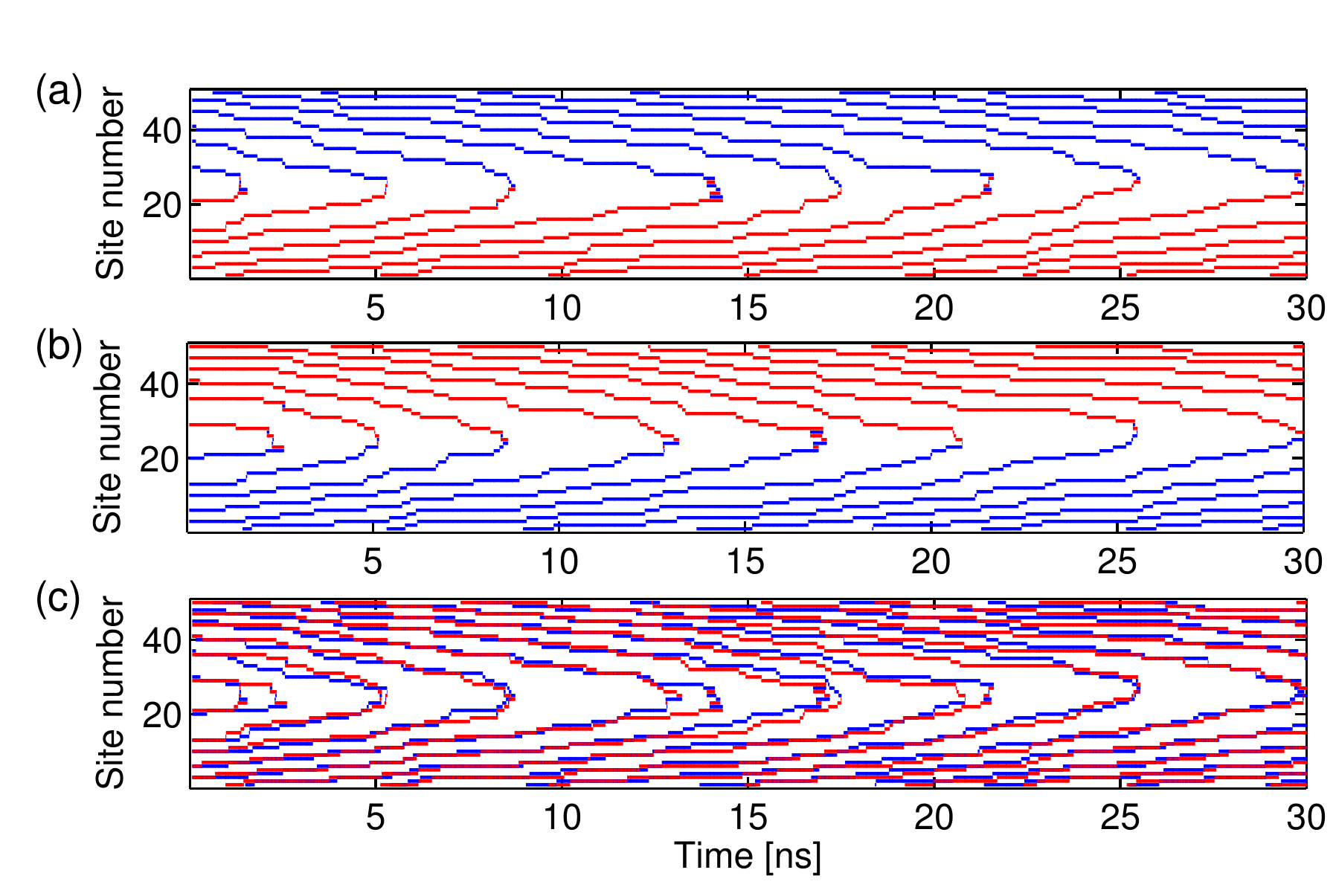}
\caption{ (Color online) Charge occupancy diagrams for escalator bias $\Delta{V}=20$ mV within the (a) upper and (b) lower rails with weak $C_C$. (c) shows (a) superimposed on (b). Nearly identical correlations suggest temporally correlated currents with charge carriers consisting of very strongly bound dipole states.
}
\label{fig:charge_s-t_escalator_weak}
\end{figure}

\begin{figure}[th!]
\includegraphics[width=0.483\textwidth]{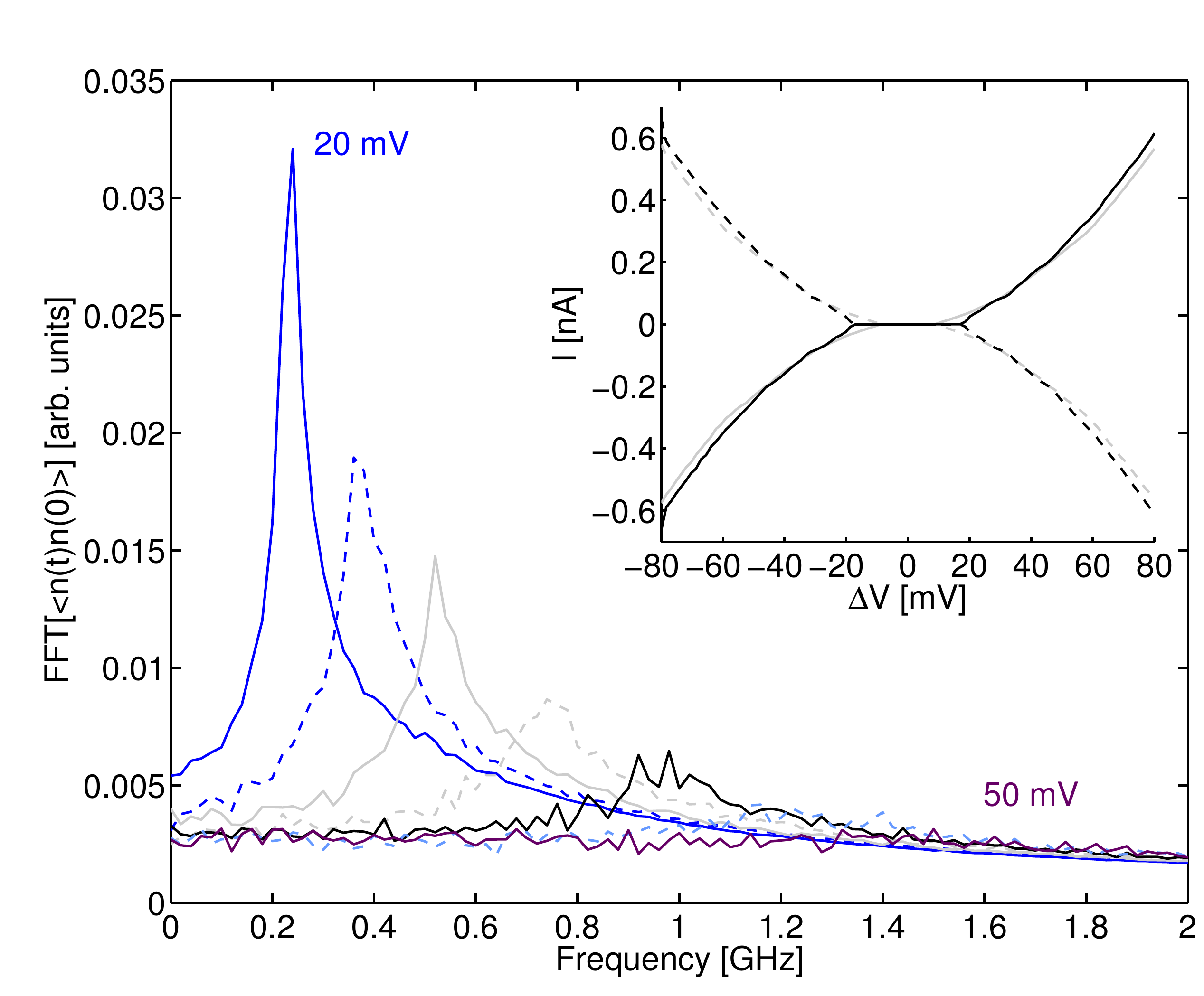}
\caption{(Color online) 
Spectral response of the charge-charge correlation function for escalator bias with weak $C_C$ calculated at $n=5$, for increasing voltage in steps of 5 mV. Inset: Current-voltage characteristics of the upper (solid) and lower (dashed) rails measured at the first junction. Strong $C_C$ (black) induces slightly larger currents and Coulomb gaps than weak $C_C$ (gray).
}
\label{fig:corr_escalator}
\end{figure}
\begin{figure}[th!]
\centering
\includegraphics[width=0.47\textwidth]{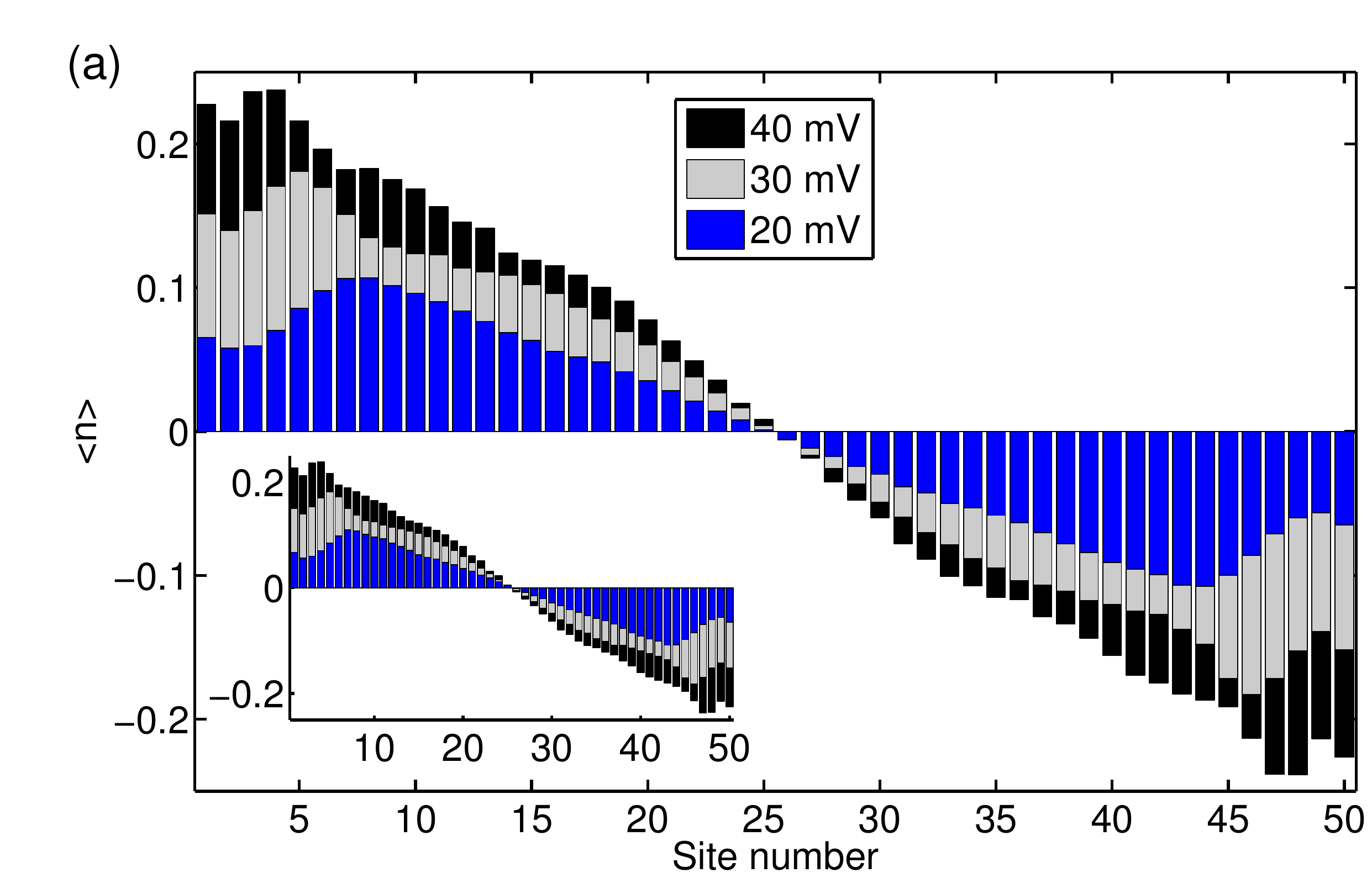}\label{fig:SDRB_weak}
\includegraphics[width=0.47\textwidth]{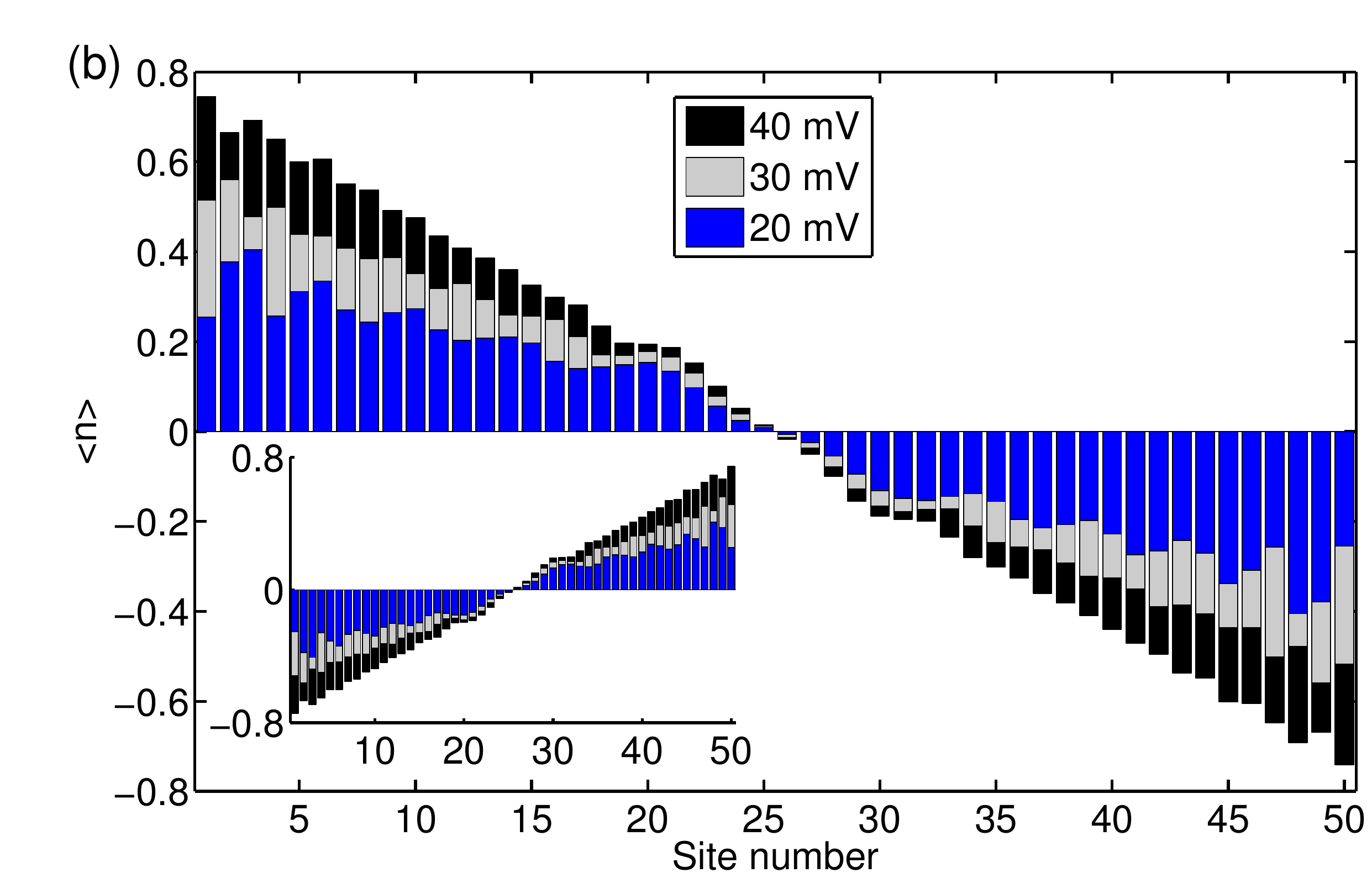}\label{fig:EB_weak_upper}
\caption{(Color online) Average charge distribution per site in the upper rails for (a) symmetric dual-rail and (b) escalator biasing  (for weak $C_C$). Insets: Lower rail average charge distribution.
}
\label{fig:bar_weak}
\end{figure}
 
For the specific case of symmetric bias, Eq.~\ref{thresv} simplifies to
\begin{equation}
\Delta{V_{\mathrm{(th)}}=\frac{q_eA_{11}}{C_J(1-A_{11}-B_{11})}}
\end{equation}
which for our model gives $\Delta{V_{\mathrm{(th)}}}=13.5$ mV for weak $C_C$ and $\Delta{V_{\mathrm{(th)}}}=11.9$ mV for strong $C_C$. These thresholds compare well to the current-voltage characteristics given in  Fig.~\ref{fig:corr_symdualrail}.

We now consider an escalator bias (i.e.~antisymmetric biasing of both rails, $\Delta{V_1}=-\Delta{V_2}$). In this regime, both rails show very strong spatial and temporal charge correlations in the current carriers, see Fig.~\ref{fig:charge_s-t_escalator_weak}. The correlations in each rail are nearly identical in space and time, suggesting that the bond between each electron-hole pair (\emph{interrail} dipole states) is very strong and that they tunnel as an effective single entity through the circuit (even though we do not consider cotunneling in our model). It is considerably more energetically favorable for a dipole to tunnel as a unit than for a dipole to break up or induce excess charge. The autocorrelation functions for an escalator bias in Fig.~\ref{fig:corr_escalator} also show strong correlations between charges. In addition, the average charge distribution for both a symmetric dual-rail and escalator bias show correlations of the charges (Fig.~\ref{fig:bar_weak}). The oscillations in the charge distribution are more pronounced for escalator bias due to the strongly correlated transport, whereas for symmetric bias, recombination site drift tends to average out the charge distribution oscillations. 

We can calculate the interaction energy for the escalator bias case, where we have two charge dipoles (i.e.~four charges) interacting,
\begin{align}
U(Q_m,Q_n)=&\frac{{Q^{2}_m}}{2C_J}\Bigg(\frac{1}{\sinh\lambda_{-}}\Bigg)+\frac{{Q^{2}_n}}{2C_J}\Bigg(\frac{1}{\sinh\lambda_{-}}\Bigg)\nonumber\\
&+\frac{Q_mQ_n}{C_J}\Bigg(\frac{e^{-\lambda_{-}|m-n|}}{\sinh\lambda_{-}}\Bigg)
\end{align}
Therefore the interacting dipoles have separation $1/\lambda_{-}\approx1.53$ for strong $C_C$ and much larger charging energy when compared to an equivalent linear array ($1/2\sinh\lambda_->>1/4\sinh\lambda$). This configuration is considerably more energetically favorable than less symmetric arrangements of the four charges, resulting in strongly locked (and correlated) electron-hole pairs.

\begin{figure}[th!]
\centering
\includegraphics[width=0.48\textwidth]{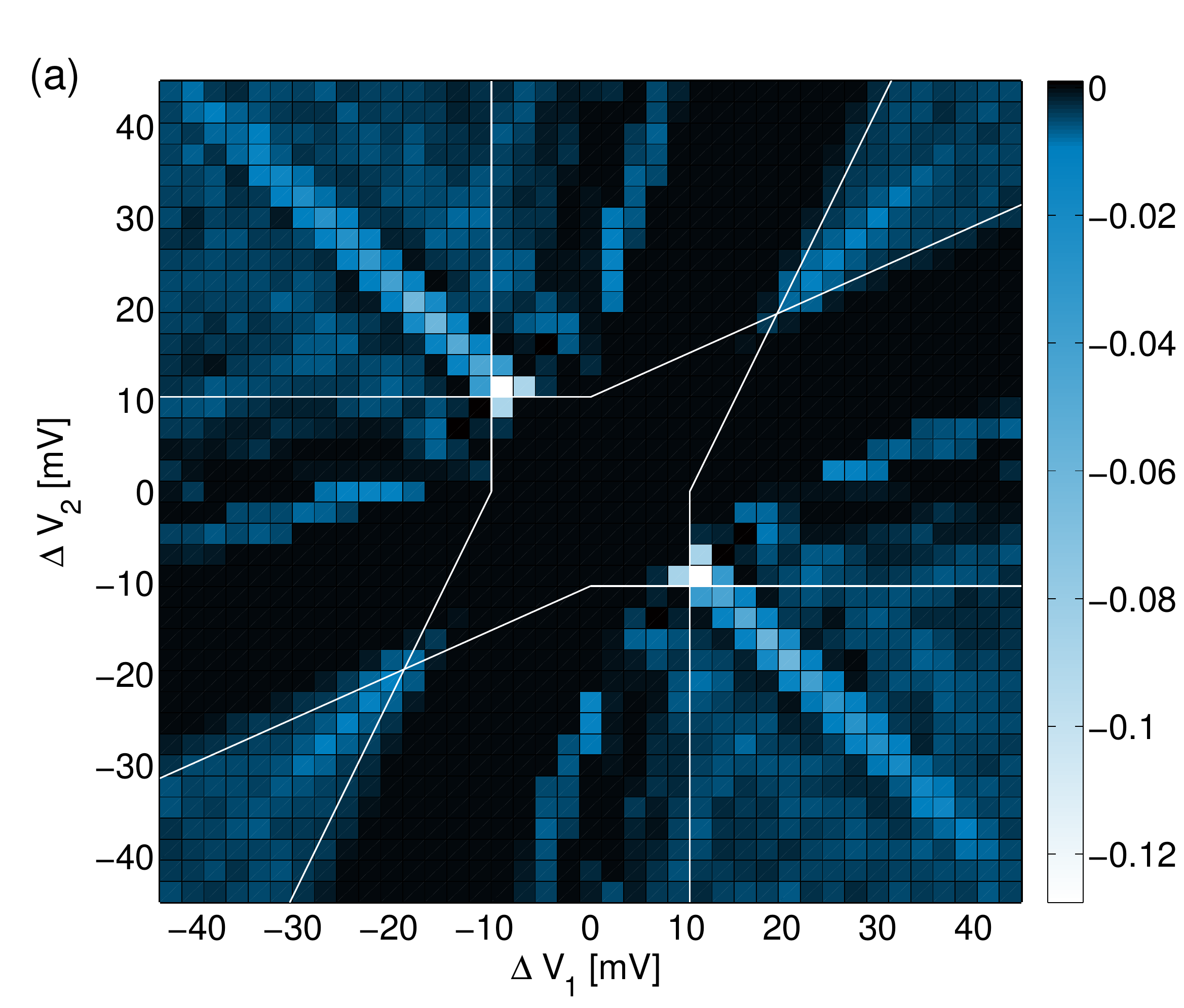}%\label{fig:corrmap_weak} %\subfloat[]{   % to include subfigure labelling.

\includegraphics[width=0.48\textwidth]{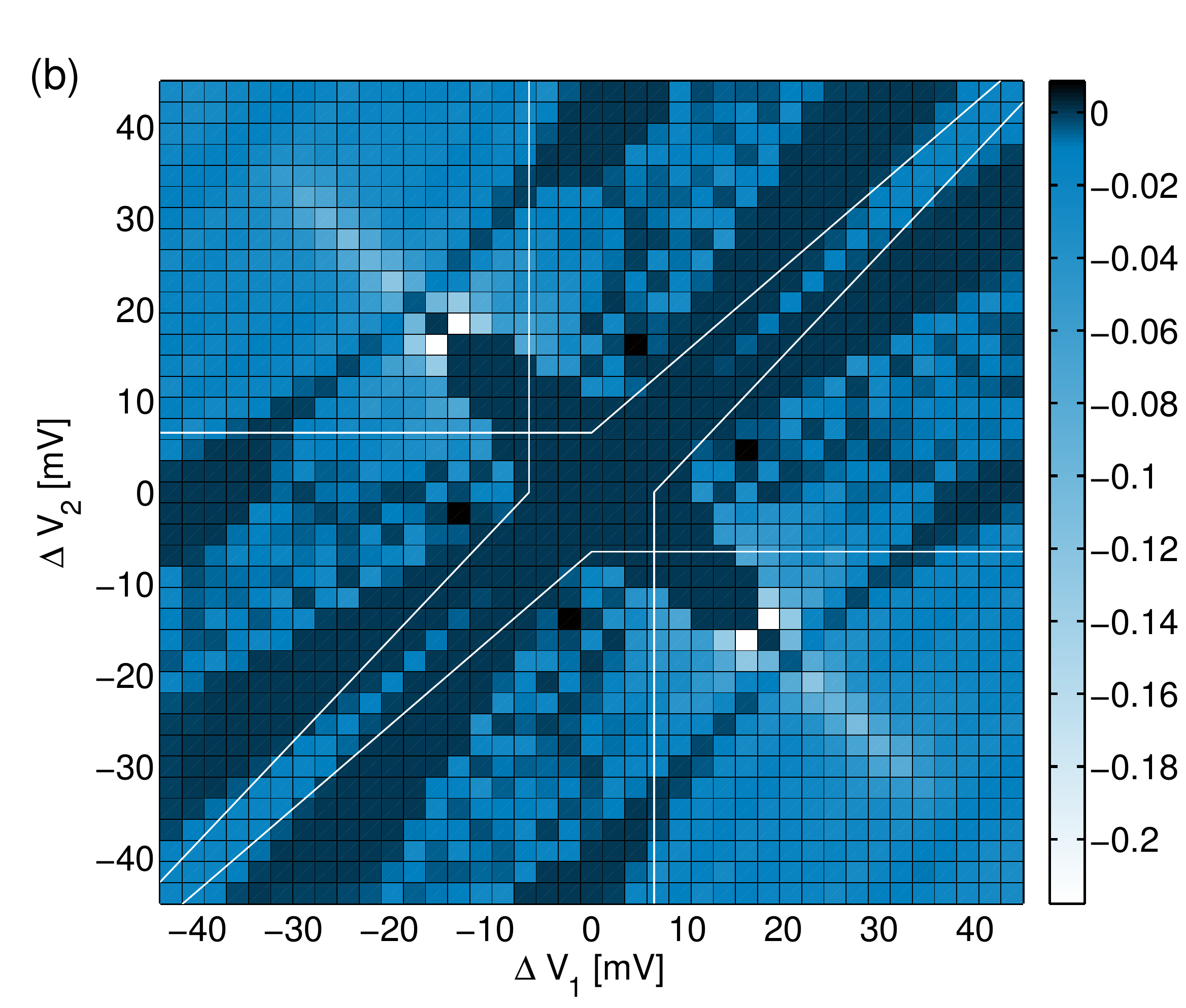}%\label{fig:corrmap_strong}
\caption{(Color online) Correlation map over the $\Delta{V_1}-\Delta{V_2}$ plane at  $n=5$ with (a) weak and (b) strong $C_C$. The currents are either uncorrelated (zero) or anticorrelated (negative). White contour lines show the analytical conduction voltage  $\Delta{V_{\mathrm{(th)}}}$.
}
\label{fig:corrmap}
\end{figure}

This creation of quasibound dipole pairs also results in a larger Coulomb gap for escalator bias, see Fig.~\ref{fig:corr_escalator} and Fig.~\ref{fig:corrmap}(b). The reduced energy required to create a dipole pair means that a larger voltage bias is required to induce flow when compared to the symmetric bias case.
%----------------------------------------------------------
%-----FLUCTUATION OF THE RECOMBINATION SITE-----
\section{\label{sec:dotrs} Fluctuation of the recombination site}
\begin{figure}[th!]
\centering
\includegraphics[width=0.233\textwidth]{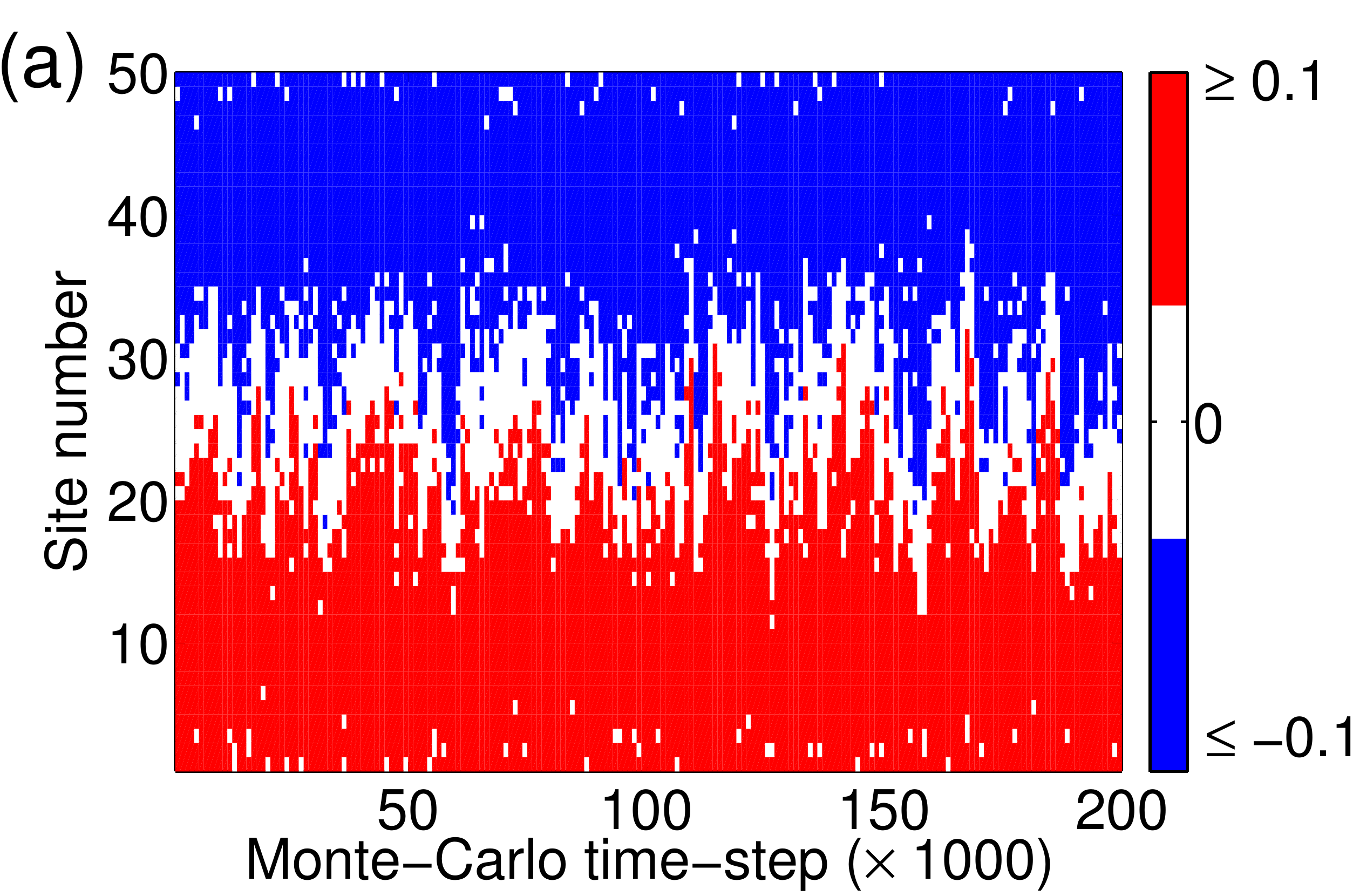}   \label{fig:pcolor_weak_upper}
\includegraphics[width=0.233\textwidth]{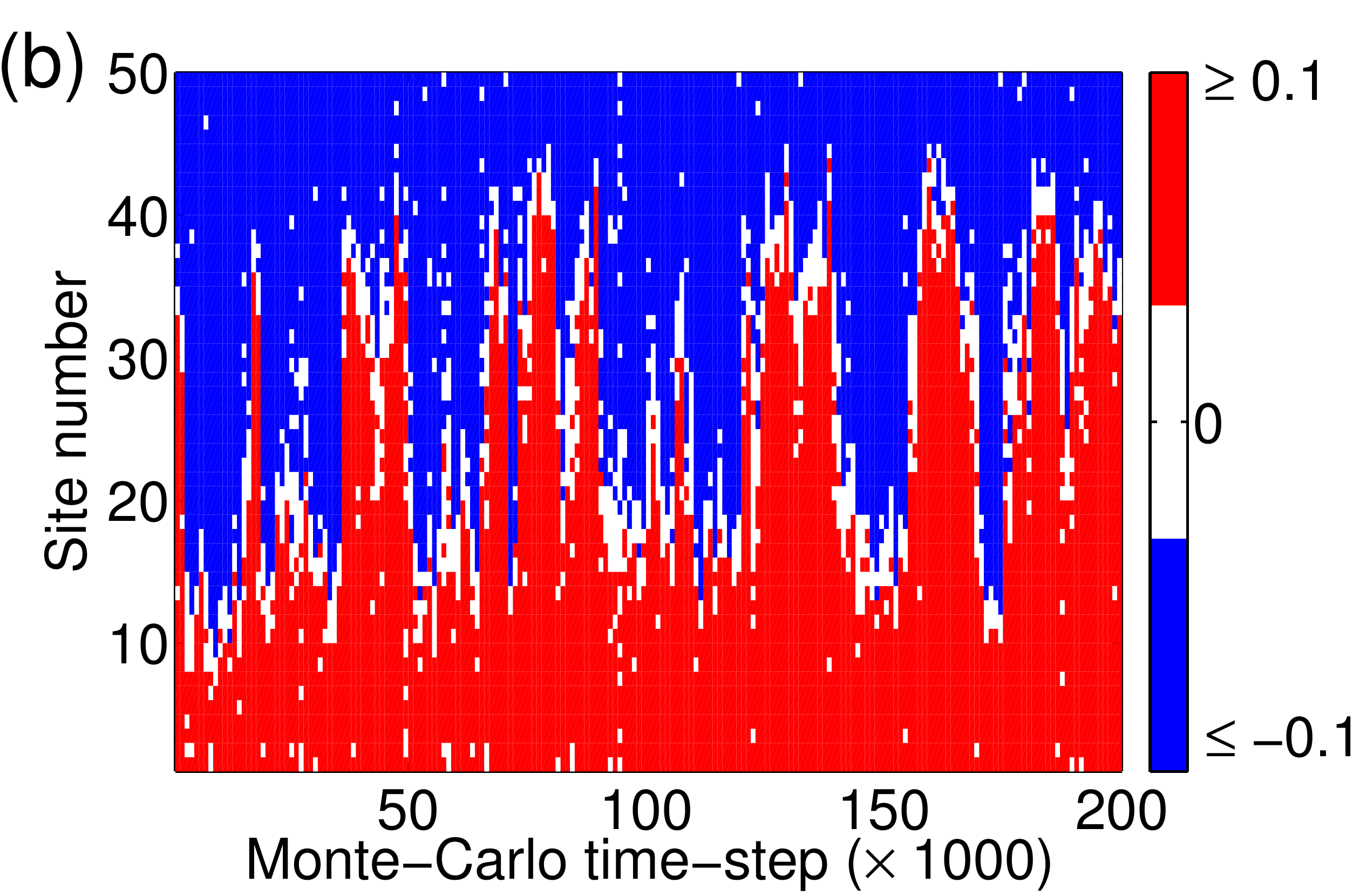}  \label{fig:pcolor_strong_upper}
\includegraphics[width=0.233\textwidth]{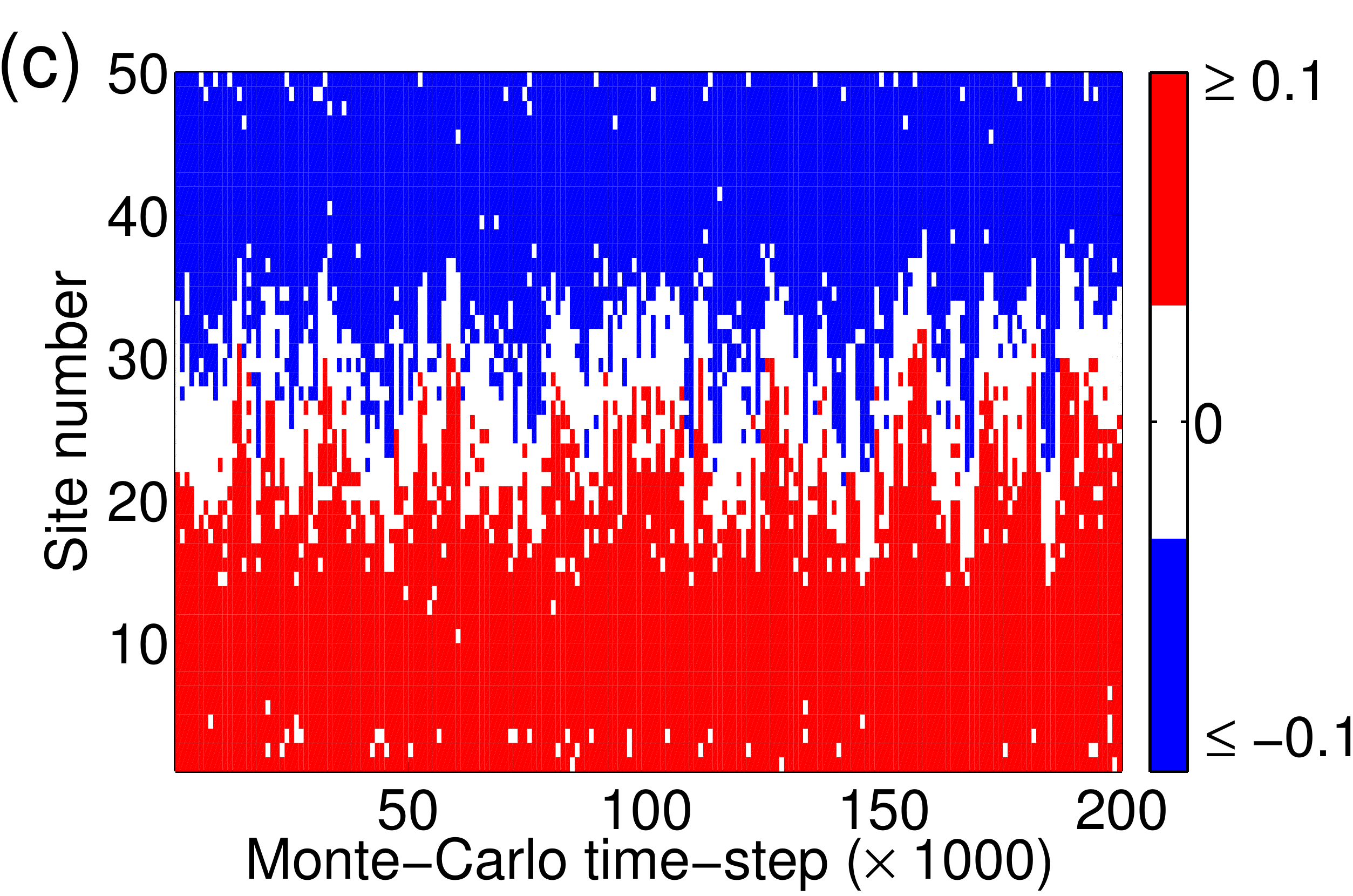}  \label{fig:pcolor_weak_lower}
\includegraphics[width=0.233\textwidth]{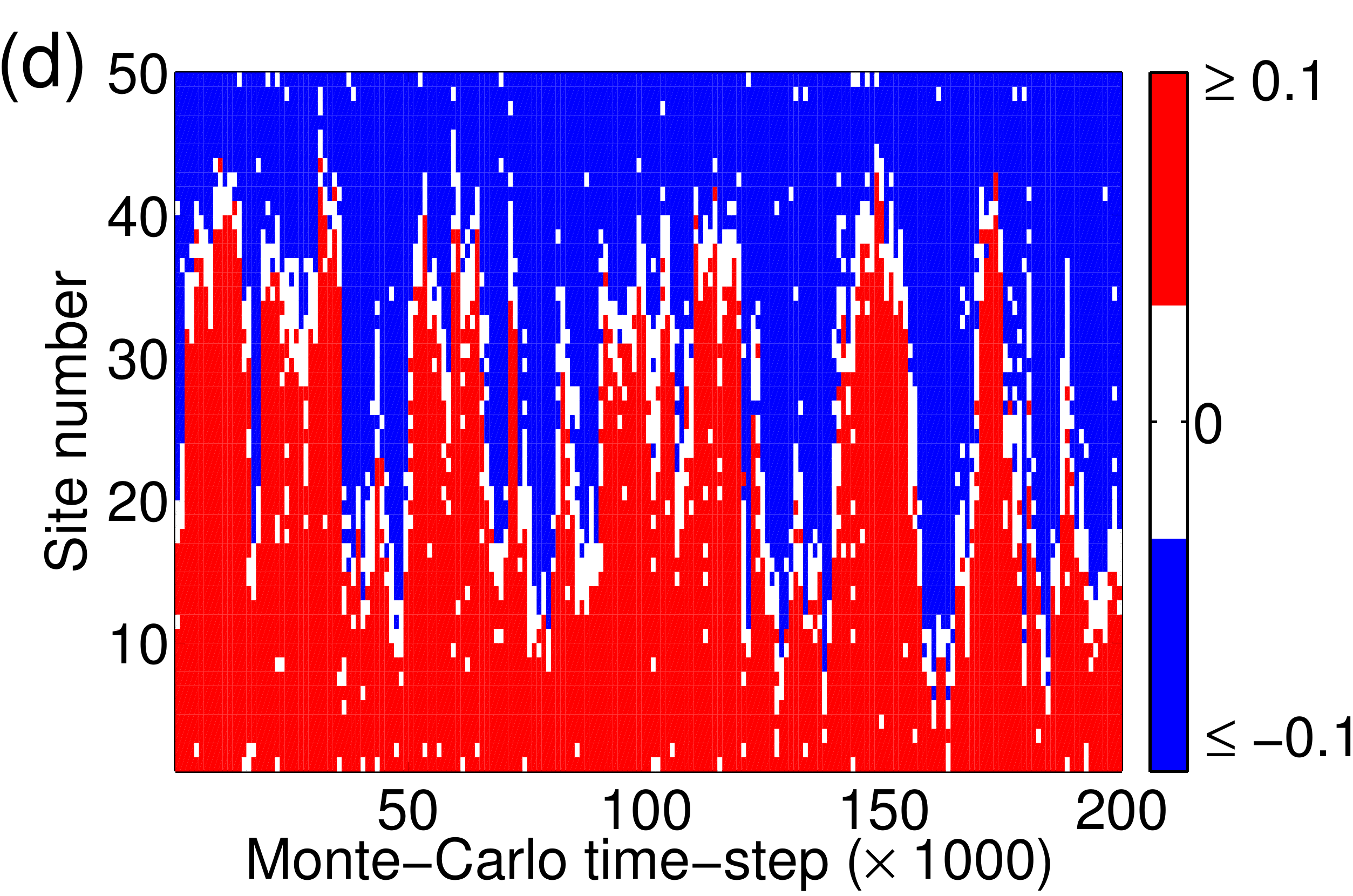}  \label{fig:pcolor_strong_lower}
\caption{ (Color online) Multiple histograms of the charge distribution plotted as a function of Monte Carlo time step for the symmetric dual-rail biased array at $\Delta{V}=20$ mV. Electrons (holes) are represented by red (blue) and the recombination site (i.e.~point of zero charge) by white. The color scale is truncated for clarity. For weak $C_C$, we see fluctuation of the opposing charge recombination site back and forth between $\sim15$ sites from either end of both the (a) upper and (c) lower rails. Strong $C_C$ causes the recombination site in both the (b) upper and (d) lower rails to fluctuate much more erratically, almost along the full length of the rails. Notice the anti-correlation of the upper and lower rail charge distributions.
}  
\label{fig:pcolor}
\end{figure}

%----------------------------------------------------------
We observed that the position within the array at which the positive and negative charge carriers recombine (the recombination site) in symmetrically biased arrays is not always the central site and is not fixed. In both the linear and bilinear arrays, the recombination site can drift several sites left or right from the center of the rail. However under certain conditions, the current recombination site can drift much more widely. While we measured drifting of the current recombination site in all biasing regimes, the effect is most prominent with an equal symmetric dual-rail bias. Fig.~\ref{fig:pcolor} shows fluctuation of the recombination site in both rails for both weak and strong $C_C$. For weak $C_C$, we see the variation of the recombination site as it drifts between approximately  $n=15$ and 35 in the upper rail. The different possible charge states of these sites all have approximately the same energy, therefore the recombination is likely to occur at any of these sites.

Strong $C_C$ causes the recombination site to fluctuate much more erratically, almost along the full length of the rails. The drifting of the recombination site is also evident in Fig.~\ref{fig:charge_s-t_symdualrail_strong} where we see each rail dominated by a particular charge carrier (i.e.~electrons or holes). In general, the recombination site in each rail is not locked and in fact we see strong anticorrelation of the entire  charge distribution. As the recombination site fluctuates  back and forth in a particular rail, the corresponding recombination site in the other rail mirrors this behavior in such a way as to guarantee an effective net charge of zero for the entire circuit. This is due to the interplay between the injection of electrons and holes in the individual rails and that of the tendency to form electron-hole pairs \emph{between} rails.

\begin{figure}[th!]
\centering
\includegraphics[width=0.47\textwidth]{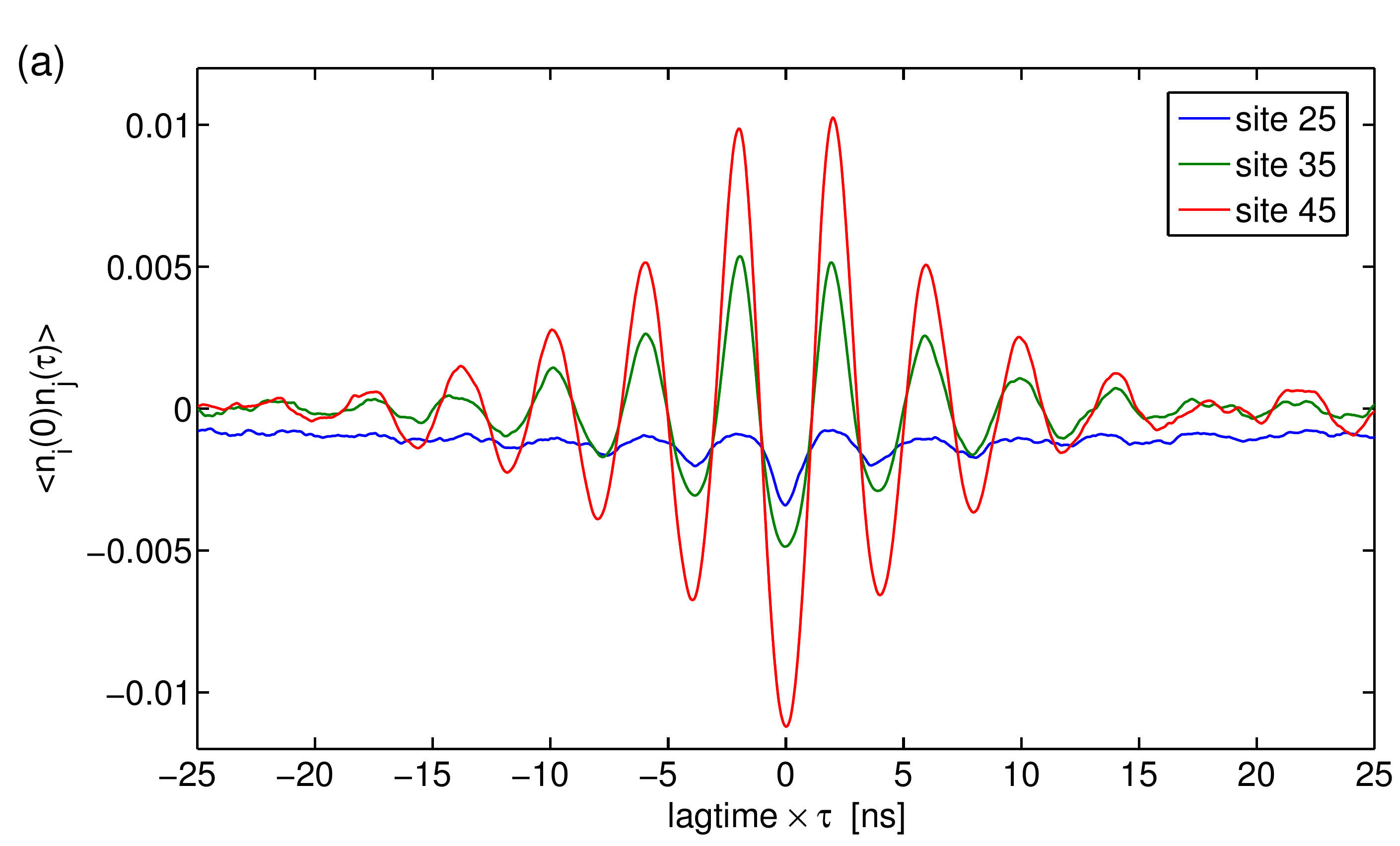}%\label{fig:crosscorr_weakDRSB}

\includegraphics[width=0.47\textwidth]{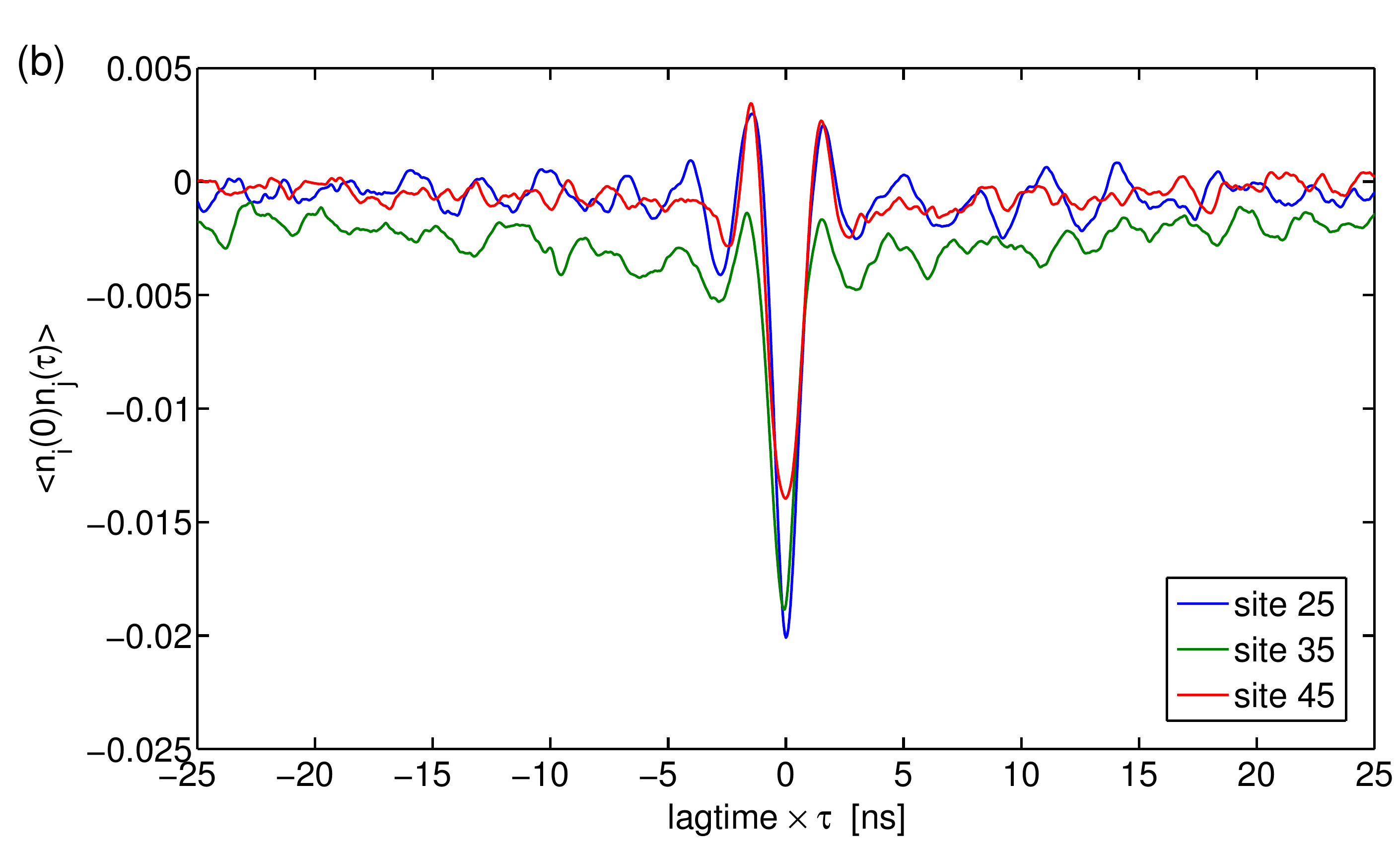}%\label{fig:crosscorr_strongDRSB}

\includegraphics[width=0.47\textwidth]{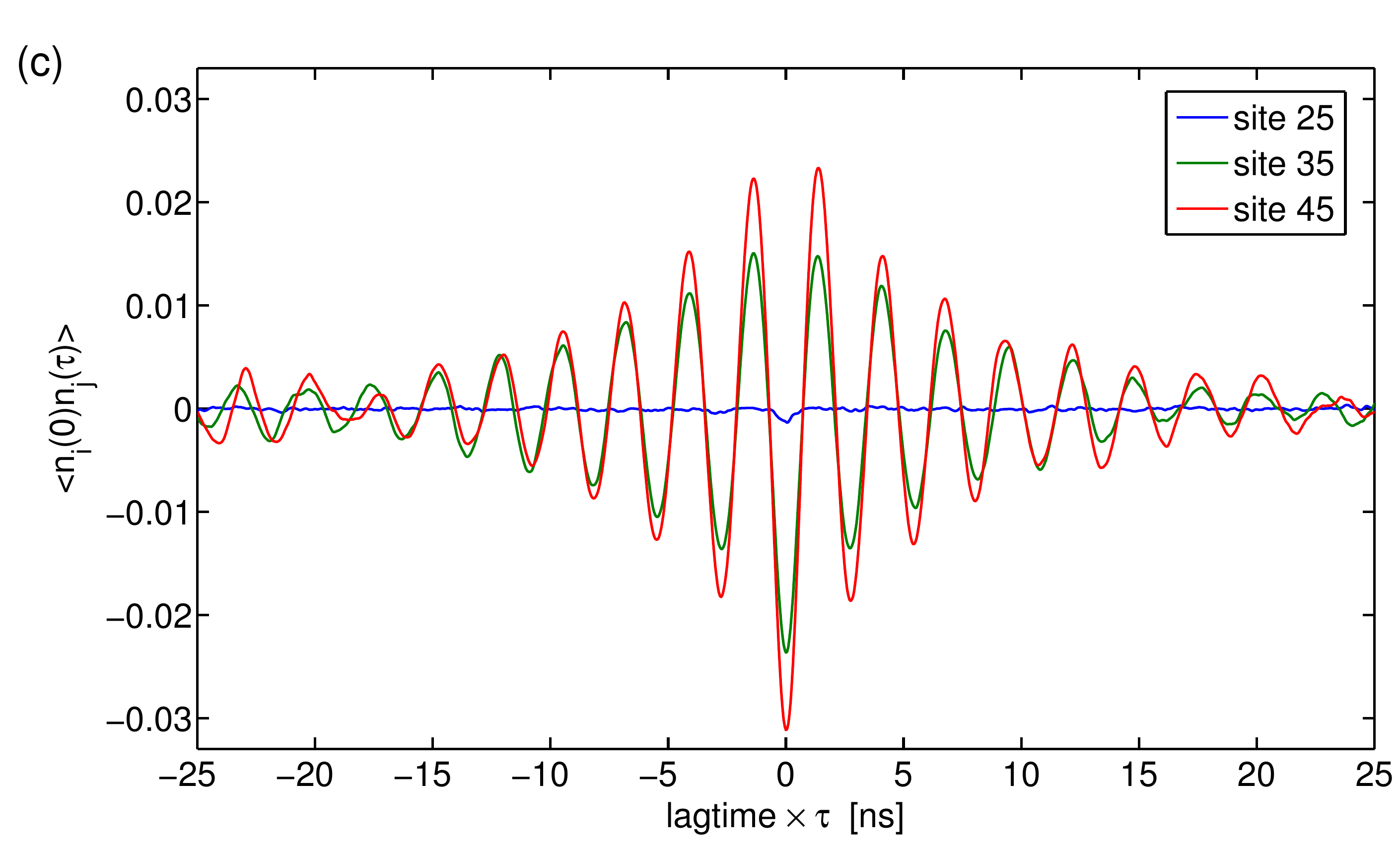}%\label{fig:crosscorr_weakEB}

\includegraphics[width=0.47\textwidth]{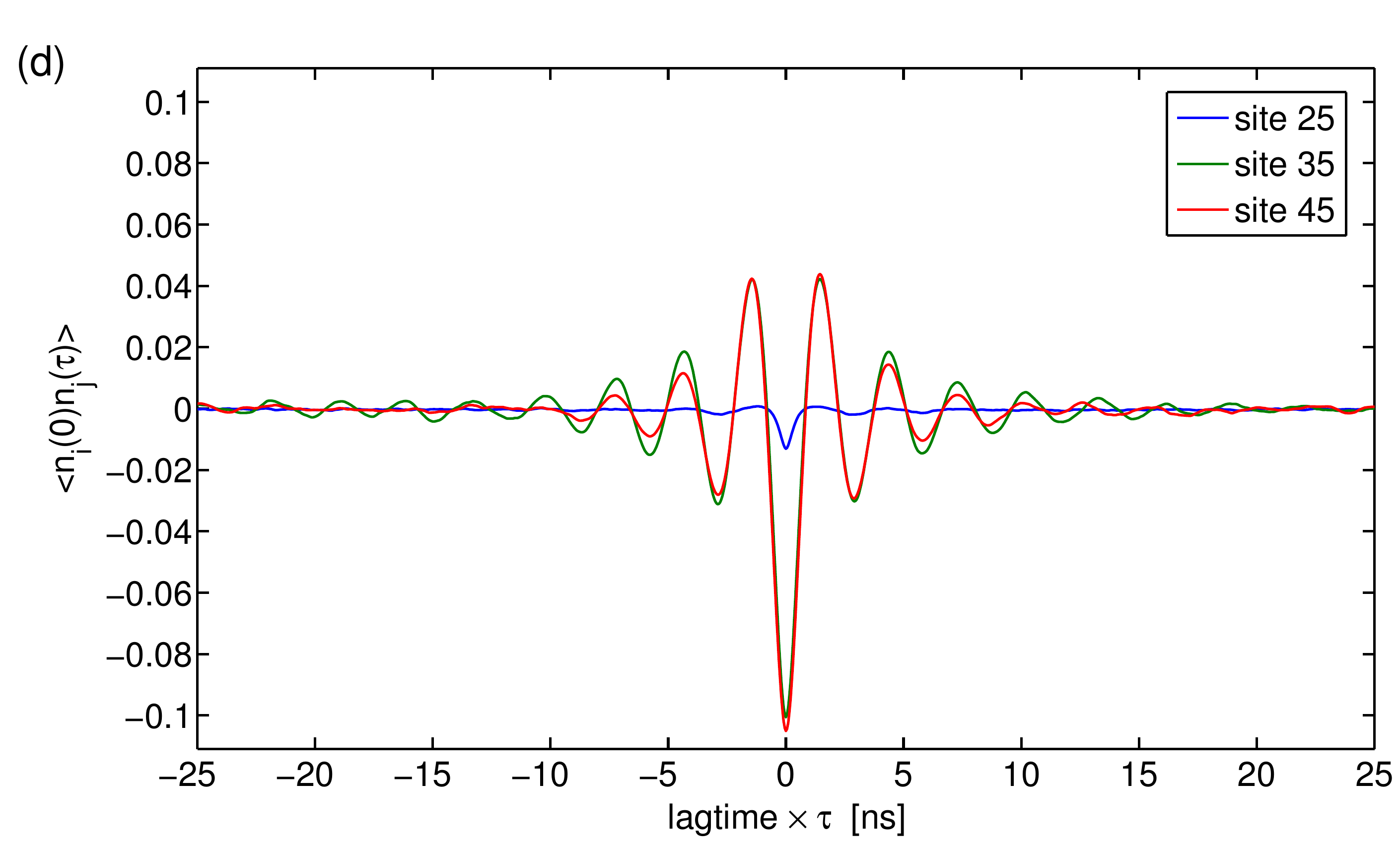}%\label{fig:crosscorr_strongEB}
\caption{(Color online) Cross-correlation functions calculated at $\Delta{V}=25$ mV in the center of the array ($n=25$) and the outer edge, $n=35$ and $n=45$ for symmetric dual-rail bias with (a) weak and (b) strong $C_C$ and escalator bias with (c) weak and (d) strong $C_C$.}
\label{fig:crosscorr} 
\end{figure}

%----------------------------------------------------------
%----- ANTICORRELATED CHARGE TRANSPORT-----
\section{\label{sec:anticorr}Anticorrelated charge transport}
%----------------------------------------------------------
We now consider the temporal correlations of charges between rails, $\langle Q_{m}(0)Q_{n}(\tau)\rangle$. This is a measure of the anticorrelation of the pairs, i.e.~correlation \emph{between} rails. This enables us to determine when the charges and therefore currents are correlated or anticorrelated.

We see from the cross-correlation functions in Fig.~\ref{fig:crosscorr}(a) that a symmetric dual-rail biased array with weak $C_C$ exhibits strong correlations between rails towards the ends of the array. However the functions are weaker towards the center as a result of the slight recombination site drift. These observations are consistent with the synchronized correlations in the outer edges of the array and slight drift of the recombination site in Fig.~\ref{fig:charge_s-t_symdualrail_weak}(c). Note that at $\tau=0$, the functions are negative at all three positions and the functions have the same period, which shows that the entire upper and lower rail charges are anticorrelated.

Due to the extreme drifting of the recombination site in the strong $C_C$ case, there is significant loss of cross correlation, see Fig.~\ref{fig:crosscorr}(b). There is however, an overall anticorrelation between rails which increases towards the center of the array. This result is a direct consequence of the anticorrelation behavior of the recombination site drift.

When an escalator bias is applied with weak $C_C$, the charges between rails are strongly locked together with only a slight drift of the recombination site, see Fig.~\ref{fig:charge_s-t_escalator_weak}(c). This is also apparent in the cross-correlation functions in Fig.~\ref{fig:crosscorr}(c), where the charges in the two rails are strongly correlated at the edges, but weaker at the center. This is an important point for experiments in which a SET is used to measure current correlations through a symmetrically biased array. Placing the SET in the middle of the array would result in weak current correlation measurements that are not indicative of those in the entire array. This point of minimum correlation can also be modified by applying an asymmetric bias ($V_1\neq-U_1$), such that the charge state of the array is either electron or hole dominated.

An escalator bias with strong $C_C$ produces an anticorrelation between rails considerably stronger than that produced in the symmetric dual-rail, strong $C_C$ case, again due to the locking of effective dipole states.

In Fig.~\ref{fig:corrmap}, we investigate cross correlation in more depth by plotting a correlation map as a function of applied voltage. This map is calculated for the cross correlations of charges at $n=5$, at zero time lag $\tau$ [i.e.~$\langle{Q_m}(0)Q_n(0)\rangle$]. As we saw in the cross-correlation functions (Fig.~\ref{fig:crosscorr}), the charges are either uncorrelated or anticorrelated. For weak $C_C$, charges are most strongly anticorrelated for an escalator bias (sweeping top left to bottom right), which is also consistent with the cross-correlation functions (Fig.~\ref{fig:crosscorr}). The strength of the correlations decreases with increasing voltage. The anticorrelation peaks in the lower-left and upper-right corners correspond to a symmetric dual-rail bias.  The solid white lines show the analytical threshold voltages for the upper and lower rails, calculated by Eq.~\ref{thresv}. The peaks within the Coulomb gap represent anticorrelations between static states in the undriven rail and current in the driven rail.

For strong $C_C$, we see that charges are again strongly anticorrelated for an escalator bias (sweeping top left to bottom right). There are a greater number of peaks within the Coulomb gap corresponding to anticorrelations between static states in the undriven rail and current in the driven rail. For strong $C_C$, the injection of dipole states in the escalator biasing regime has a larger Coulomb gap, as previously seen in  Fig.~\ref{fig:corr_escalator}. In this case, a hole cannot tunnel through one rail unless its matching electron is also injected in the other rail.
%----------------------------------------------------------
%---- CONCLUSION-----
\section{\label{sec:concl}Conclusion}
%----------------------------------------------------------
This work focuses on the nature of the correlations in space and time of the current in a biased bilinear array of nonsuperconducting tunnel junctions. We have demonstrated that both high $\Delta V$ and strong interrail capacitance destroy charge correlations within a rail. When only one rail is biased, the undriven rail does not show temporal correlated charge transport, however static quasiparticle states are created which show some spatial correlation. When both rails are biased we observe temporally and spatially synchronized correlations between rails . Furthermore, both an escalator and symmetric dual-rail bias induce anticorrelated currents. We also observed significant drifting of the recombination site in a symmetrically biased array.
%----------------------------------------------------------
%----- ACKNOWLEDGEMENTS-----
%----------------------------------------------------------
\begin{acknowledgments}
We acknowledge helpful discussions with N. Vogt, A. Greentree, T. Duty and P. Delsing. This work was supported by the Victorian Partnership for Advanced Computing (VPAC).
\end{acknowledgments}
%----------------------------------------------------------
%----- REFERENCES-----
%----------------------------------------------------------
\bibliography{references_paper1}

\end{document}